\documentclass[fractalfract,article,accept,pdftex,oneauthor]{Definitions/mdpi} 

\graphicspath{ {./images/} }
\usepackage{wrapfig}
\usepackage{enumitem}
\usepackage{amsmath,stackengine}
\stackMath
\usepackage{amssymb}
\usepackage{marginnote}
\usepackage{comment}
\usepackage{url}
\usepackage{esint}
\usepackage{multirow}
\usepackage{caption}
\usepackage{tikz}
\usetikzlibrary{arrows.meta}

\newcommand{\e}[1]{^{[#1]}}

\usepackage{mathtools}
\usepackage{listings}
\usepackage{xcolor}
\usepackage{siunitx}
\usepackage{pgfplots}
\pgfplotsset{compat=1.9, , every axis/.append style={font=\scriptsize}}
\usepackage{xspace}
\usepackage{algorithm}
\usepackage{algpseudocode}

\newcommand{\mb}{\mathbf}

\firstpage{1} 
\pubvolume{9}
\issuenum{9}
\articlenumber{584}
\pubyear{2025}
\copyrightyear{2025}
\externaleditor{Carlo Cattani and Christoph Bandt} 
\datereceived{9 July 2025} 
\daterevised{15 August 2025} 
\dateaccepted{1 September 2025} 
\datepublished{3 September 2025} 
\hreflink{https://doi.org/10.3390/\linebreak fractalfract9090584} 
\doinum{10.3390/fractalfract9090584}

\Title{Chaotic Dynamics and Fractal Geometry in Ring Lattice Systems of Nonchaotic Rulkov Neurons}

\TitleCitation{Chaotic Dynamics and Fractal Geometry in Ring Lattice Systems of Nonchaotic Rulkov Neurons}

\Author{Brandon B. Le \orcidA{}}

\AuthorNames{Brandon B. Le}

\isAPAStyle{%
       \AuthorCitation{Le, B.B.}
         }{%
        \isChicagoStyle{%
        \AuthorCitation{Le, Brandon B.}
        }{
        \AuthorCitation{Le, B.B.}
        }
}

\address{%
Department of Physics, University of Virginia, Charlottesville, VA 22904-4714, USA; sxh3qf@virginia.edu}

\abstract{This paper investigates the complex dynamics and fractal attractors that arise in a 60-dimensional ring lattice system of electrically coupled nonchaotic Rulkov neurons. While networks of chaotic Rulkov neurons have been widely studied, systems of nonchaotic Rulkov neurons have not been extensively explored due to the piecewise complexity of the nonchaotic Rulkov map. Here, we find that rich dynamics emerge from the electrical coupling of regular-spiking Rulkov neurons, including chaotic spiking, synchronized chaotic bursting, and synchronized hyperchaos. By systematically varying the electrical coupling strength between neurons, we also uncover general trends in the maximal Lyapunov exponent across the system's dynamical regimes. By means of the Kaplan--Yorke conjecture, we examine the fractal geometry of the ring system’s high-dimensional chaotic attractors and find that these attractors can occupy as many as 45 of the 60 dimensions of state space. We further explore how variations in chaotic behavior---quantified by the full Lyapunov spectra---correspond to changes in the attractors’ fractal dimensions. This analysis advances our understanding of how complex collective behavior can emerge from the interaction of multiple simple neuron models and highlights the deep interplay between dynamics and geometry in high-dimensional systems.}


\begin{document}



\section{Introduction}

Biological neurons are well known to exhibit a wide variety of interesting dynamic behaviors, including nonchaotic and chaotic spiking and bursting~\cite{izhikevich-article}. Since the pioneering work of Hodgkin and Huxley~\cite{hh}, many continuous-time neuron models have been developed in an attempt to model the complex behavior of biological neurons~\cite{chay, buchholtz, izhikevich-model, fh}. In~order to capture the dynamics of neurons with fast bursts of spikes on top of slow oscillations, many of these models are slow--fast dynamical systems~\cite{hindmarsh, rinzel, izh-map, izhikevich-model, courbage, omelchenko}. However, these systems of nonlinear differential equations are often unwieldy to work with, posing a significant computational obstacle in modeling the behavior of many-neuron systems~\cite{izh-time}. As~a result, some discrete-time neuron models have been proposed, including Rulkov's simple two-dimensional slow--fast models~\cite{rulkov, rulkov2}.

These models, often called  chaotic and nonchaotic Rulkov models~\cite{ibarz}, are capable of modeling both chaotic and nonchaotic spiking and bursting behaviors, and~they are computationally efficient, allowing for the study of neuron systems with a complex architecture. The~chaotic Rulkov model has been well studied in the literature~\cite{ibarz, vries, luo, min, bao, dePontes}, but~in this paper, we will focus on the nonchaotic Rulkov model, which also produces rich and interesting dynamics. As~expected, the~most direct application of the (nonchaotic) Rulkov map is in modeling neuronal dynamics~\cite{rulkov}, but~it has also shown application in stability analysis~\cite{wang2}, control of chaos~\cite{lopez}, symbolic analysis~\cite{budzinski}, final-state sensitivity~\cite{le}, machine learning~\cite{ge}, information patterns~\cite{njitacke}, and~digital watermarking~\cite{ding}. Therefore, it is a worthwhile system to study purely due to its dynamical and geometrical~properties.

In this paper, we are interested in lattice systems of coupled nonchaotic Rulkov neurons. In existing research, networks of coupled chaotic Rulkov neurons have received much attention, especially regarding the synchronization of chaotic Rulkov neuron networks. For~example, existing studies include two chaotic Rulkov neurons coupled with chemical synapses~\cite{hu}, two chaotic Rulkov neurons with a chemical synaptic and inner linking coupling~\cite{rakshit}, the~complete synchronization of an electrically coupled chaotic Rulkov neuron network~\cite{sun}, synchronization in a network of chaotic Rulkov neurons with a leader--follower structure \cite{marghoti}, 
and~coupling a discrete memristor into a chaotic Rulkov neuron~\cite{mem-1, mem-2, mem-3, mem-4, mem-5}. However, coupled systems of nonchaotic Rulkov neurons have not received nearly as much attention due to the complexity of the piecewise function $f$ present in the nonchaotic Rulkov map (Equation~\eqref{eq:rulkov_1_fast_equation}).

In this paper, we investigate neurons arranged in a ring lattice, which is a common topology used when studying coupled dynamical systems~\cite{banerjee, jampa, chen}. Specifically, we are interested in a ring of $\zeta$ electrically coupled nonchaotic Rulkov neurons $\mathbf{x}_0,\mathbf{x}_1,\hdots,\mathbf{x}_{\zeta-1}$, each with a flow of current with its neighbors (see Figure~\ref{fig:ring-30-rulkov-1-neurons}). Osipov~et~al. \cite{osipov} qualitatively describe the dynamics of a similar Rulkov ring lattice system, noting the emergence of complex dynamics from Rulkov neurons in the nonchaotic spiking regime. Building on this previous work, this paper involves a quantitative, numerical analysis of the chaotic dynamics emerging from three different regimes of the ring lattice system, each with different individual neuron behaviors. The~piecewise function $f$ present in the iteration function of each neuron in the ring is found to yield an impressively complex Jacobian matrix. Using this, the~dynamics of this system are explored with greater generality over a wide range of electrical coupling strength values through numerical simulation and computation of the system's maximal Lyapunov exponents. The~main focus of this work is to analyze the fractal geometry of the system's high-dimensional chaotic attractors and how it changes as the electrical coupling strength varies. In~particular, we explore the complex relationship between the chaotic trajectories that the system follows on these attractors and the geometric structure and complexity of the attractors~themselves.

This paper is organized as follows. Section~\ref{sec:dynamics} describes the model and the three regimes of interest, then presents the qualitative and quantitative analysis of their complex dynamics. Section~\ref{sec:geometry} overviews the Kaplan--Yorke conjecture and uses it to approximate the fractal dimensions of the system's attractors in 60-dimensional state space. Finally, Section~\ref{sec:conclusions} summarizes our results, discusses their implications, and provides suggestions for future~research.

\begin{figure}[H]
    \begin{tikzpicture}[scale=1]
        \draw[yellow, very thick] (0, 0) circle [radius=3cm];
        \foreach \a in {0,1,...,29}{
            \filldraw[blue] (\a*360/30: 3cm) circle [radius=2.5pt];
            \draw (\a*360/30+96: 3.4cm) node{$\mathbf{x}_{\a}$};
        }
    \end{tikzpicture}
    \caption{Visualization of a ring of $\zeta=30$ Rulkov neurons. Neurons are shown as blue points, and~electrical coupling connections are shown in~gold.}
    \label{fig:ring-30-rulkov-1-neurons}
\end{figure}


\section{The Model and~its Dynamics}
\label{sec:dynamics}

The nonchaotic Rulkov map is defined by the following iteration function:
\begin{equation}
    \begin{pmatrix}
        x_{k+1} \\
        y_{k+1}
    \end{pmatrix}
    =
    \begin{pmatrix}
        f(x_k, y_k; \alpha) \\
        y_k - \mu(x_k - \sigma)
    \end{pmatrix},
    \label{eq:rulkov-map}
\end{equation}
where $f$ is the piecewise function
\begin{equation}
    f(x,y;\alpha) = 
    \begin{cases}
        \alpha/(1-x) + y, & x\leq 0 \\
        \alpha + y, & 0 < x < \alpha + y \\
        -1, & x\geq \alpha + y
    \end{cases}.
    \label{eq:rulkov_1_fast_equation}
\end{equation}
Here, $\mb{x}_k = (x_k, y_k)$ is the state of the system at time step $t=k$, $x$ is the fast variable representing the voltage of the neuron, $y$ is the slow variable, and~$\alpha$, $\sigma$, and~$\mu$ are parameters. In the original paper that introduces the Rulkov map~\cite{rulkov}, the parameter $\sigma'=\sigma+1$ is used, but~we use the slightly modified form from Ref.~\cite{ibarz}. To~make $y$ a slow variable, $0<\mu\ll1$ is needed, so we choose the standard value of $\mu=0.001$. To~understand the role of the parameters $\sigma$ and $\alpha$, we first observe the effect of $y$ on the fast-variable map $f$, namely, that increasing $y$ raises the height of $f$, which results in a quicker increase in $x$ before the resetting mechanism (the third piece of $f$) is reached. In~other words, a~higher $y$ results in faster spikes. From~the slow-variable iteration function, it is clear that $\sigma$ controls the value of $x$, which keeps $y$ constant, and~if the average value of $x$ is less than $\sigma$, then $y$ will increase until the average value of $x$ reaches $\sigma$, and~vice~versa. Therefore, $\sigma$ is an ``excitation parameter,'' since a higher value of $\sigma$ will cause $y$ to increase, increasing the frequency of spikes. The~role of the parameter $\alpha$ is more subtle, but~its main purpose is to control the existence of a stable point and the bursting regime, or~oscillations between spiking and silence. Specifically, for~$\alpha>4$, certain values of $\sigma$ will result in bursting behavior. For~a more detailed explanation of the behavior of individual nonchaotic Rulkov neurons and the roles of the parameters $\alpha$ and $\sigma$, see Refs.~\cite{rulkov, bn, ramirez}.

In experiments, biologists can alter the behavior of biological neurons by injecting the cell with a direct electrical current through an electrode~\cite{rulkov}. Modeling an injection of current from a DC voltage source requires a slight modification to the Rulkov iteration function given in Equation~\eqref{eq:rulkov-map}:
\begin{equation}
    \begin{pmatrix}
        x_{k+1} \\
        y_{k+1}
    \end{pmatrix}
    =
    \begin{pmatrix}
        f(x_k,y_k + \beta_k;\alpha) \\
        y_k - \mu(x_k - \sigma_k)
    \end{pmatrix},
    \label{eq:rulkov-map-injected-current}
\end{equation}
where the parameters $\beta_k$ and $\sigma_k$ model a time-varying injected current. Here, we are interested in coupling Rulkov neurons with a flow of current. To~model this, say we have some coupled Rulkov neurons with states $\mathbf{x}_i$, where $i$ denotes the neuron index. Then, mirroring Equation~\eqref{eq:rulkov-map-injected-current}, the~iteration function of the $i$th coupled neuron is defined as
\begin{equation}
    \begin{pmatrix}
        x_{i,k+1} \\
        y_{i,k+1}
    \end{pmatrix} = 
    \begin{pmatrix}
        f(x_{i,k}, y_{i,k} + \mathfrak{C}_{i,x}(k);\alpha_i) \\[2px]
        y_{i,k} - \mu x_{i,k} + \mu[\sigma_i + \mathfrak{C}_{i,y}(k)]
    \end{pmatrix},
    \label{eq:rulkov_coupled_mapping}
\end{equation}
where $\mathbf{x}_{i,k}$ is the state of the neuron $\mathbf{x}_i$ at the time step $k$. The~coupling parameters $\mathfrak{C}_{i,x}(t)$ and $\mathfrak{C}_{i,y}(t)$ depend on the structural arrangement of the system's neurons in physical space, as~well as the electrical coupling strength (or coupling conductance) $g$ between the~neurons. 

In electrically coupled neuron systems, the~difference in the voltages, or~fast variables, of~two adjacent neurons is what results in a flow of current between them. For~this reason, we model the electrical coupling parameters $\mathfrak{C}_{i,x}(t)$ and $\mathfrak{C}_{i,y}(t)$ to be proportional to the difference between the voltage of a given neuron $\mathbf{x}_i$ and the voltages of its adjacent neurons $\mathbf{x}_j$. Specifically, the~electrical coupling parameters of the neuron $\mathbf{x}_i$ are defined as
\begin{align}
    \mathfrak{C}_{i,x}(t) = \frac{\beta^c_i}{|\mathcal{N}_i|} \sum_{j\in\mathcal{N}_i} g_{ji}(x_{j,t}-x_{i,t}), \label{eq:electrical_coupling_parameter_general_x} \\
    \mathfrak{C}_{i,y}(t) = \frac{\sigma^c_i}{|\mathcal{N}_i|} \sum_{j\in\mathcal{N}_i} g_{ji}(x_{j,t}-x_{i,t}), \label{eq:electrical_coupling_parameter_general_y}
\end{align}
where $\mathcal{N}_i$ is the set of neurons that are adjacent to $\mathbf{x}_i$, and $g_{ji}$ is the electrical coupling strength from $\mathbf{x}_j$ to $\mathbf{x}_i$ \cite{ibarz}.

The model investigated in this paper is a ring lattice of $\zeta$ electrically coupled nonchaotic Rulkov neurons. This lattice structure is visualized in Figure~\ref{fig:ring-30-rulkov-1-neurons} for $\zeta=30$, where neurons are represented by blue points and the electric coupling connections are shown in gold. To~determine the coupling parameters for each of these neurons, let $\beta^c_i=\sigma^c_i=1$ for simplicity. We will also assume that all couplings are equivalent and symmetric: $g=g_{ji}$ for all $i\neq j$. Because~of the circular nature of this lattice system, $\mathcal{N}_i$ can be written as
\begin{equation}
    \mathcal{N}_i = \{\mathbf{x}_{(i-1)\bmod\zeta},\mathbf{x}_{(i+1)\bmod\zeta}\},
\end{equation}
which accounts for the fact that $\mathcal{N}_0 = \{\mathbf{x}_{\zeta-1},\mathbf{x}_1\}$ and $\mathcal{N}_{\zeta-1} = \{\mathbf{x}_{\zeta-2},\mathbf{x}_0\}$. Then, from\linebreak  Equations~\eqref{eq:electrical_coupling_parameter_general_x} and \eqref{eq:electrical_coupling_parameter_general_y}, the~coupling parameters of this ring system are
\begin{equation}
    \begin{split}
        \mathfrak{C}_{i} &= \mathfrak{C}_{i,x} = \mathfrak{C}_{i,y} \\
        &= \frac{g}{2}[(x_{(i-1)\bmod\zeta}-x_i)+(x_{(i+1)\bmod\zeta}-x_i)] \\
        &= \frac{g}{2}[x_{(i-1)\bmod\zeta}+x_{(i+1)\bmod\zeta}-2x_i].
    \end{split}
    \label{eq:ring-coup-params}
\end{equation}

The state vector of this entire ring system with all $\zeta$ neurons can be written as
\begin{equation}
    \mathbf{X} = \begin{pmatrix}
        X\e{1} \\
        X\e{2} \\
        X\e{3} \\
        X\e{4} \\
        \vdots \\[3px]
        X\e{2\zeta-1} \\
        X\e{2\zeta}
    \end{pmatrix} = 
    \begin{pmatrix}
        x_0 \\
        y_0 \\
        x_1 \\
        y_1 \\
        \vdots \\
        x_{\zeta-1} \\
        y_{\zeta-1}
    \end{pmatrix},
    \label{eq:ring-system-state-vector}
\end{equation}
where $X\e{p}$ is the $p$th dimension of the state vector $\mb{X}$. The~state space of this ring lattice system is $2\zeta$-dimensional, since we have one slow variable and one fast variable for each of the $\zeta$ neurons in the ring. Plugging the coupling parameters (Equation~\eqref{eq:ring-coup-params}) into the general iteration function for coupled Rulkov maps (Equation~\eqref{eq:rulkov_coupled_mapping}) for each neuron in the ring yields the $2\zeta$-dimensional iteration function for the system:
\begin{adjustwidth}{-\extralength}{0cm}
\begin{equation}
    \mathbf{F}(\mathbf{X})
    = \begin{pmatrix}
        F\e{1}(x_0,y_0,x_1,y_1,\hdots,x_{\zeta-1},y_{\zeta-1}) \\[4px]
        F\e{2}(x_0,y_0,x_1,y_1,\hdots,x_{\zeta-1},y_{\zeta-1}) \\[4px]
        F\e{3}(x_0,y_0,x_1,y_1,\hdots,x_{\zeta-1},y_{\zeta-1}) \\[4px]
        F\e{4}(x_0,y_0,x_1,y_1,\hdots,x_{\zeta-1},y_{\zeta-1}) \\[4px]
        \vdots \\[4px]
        F\e{2\zeta-1}(x_0,y_0,x_1,y_1,\hdots,x_{\zeta-1},y_{\zeta-1}) \\[4px]
        F\e{2\zeta}(x_0,y_0,x_1,y_1,\hdots,x_{\zeta-1},y_{\zeta-1})
    \end{pmatrix} = \begin{pmatrix}
        f\Big(x_0,y_0+\frac{g}{2}(x_{\zeta-1}+x_1-2x_0);\alpha_0\Big) \\[4px]
        y_0 - \mu x_0 + \mu\Big[\sigma_0 + \frac{g}{2}(x_{\zeta-1}+x_1-2x_0)\Big] \\[4px]
        f\Big(x_1,y_1+\frac{g}{2}(x_0+x_2-2x_1);\alpha_1\Big) \\[4px]
        y_1 - \mu x_1 + \mu\Big[\sigma_1 + \frac{g}{2}(x_0+x_2-2x_1)\Big] \\[4px]
        \vdots \\[4px]
        f\Big(x_{\zeta-1},y_{\zeta-1}+\frac{g}{2}(x_{\zeta-2}+x_0-2x_{\zeta-1});\alpha_{\zeta-1}\Big) \\[4px]
        y_{\zeta-1} - \mu x_{\zeta-1} + \mu\Big[\sigma_{\zeta-1} + (x_{\zeta-2}+x_0-2x_{\zeta-1})\Big] 
    \end{pmatrix}.
    \label{eq:ring-iteration-func}
\end{equation}
\end{adjustwidth}

By using numerical simulations to systematically vary $\zeta$ (see Appendix \ref{appen:pseudocode}), it can be found that for $\zeta\gtrsim 4$, varying $\zeta$ has no effect on the qualitative behavior of the ring lattice system. Therefore, we choose to perform our computational analysis on a system with the architecture shown in Figure~\ref{fig:ring-30-rulkov-1-neurons}: a ring of $\zeta = 30$ electrically coupled Rulkov neurons. This network size strikes a balance between computational tractability and dynamical richness: it is large enough to support complex collective behaviors such as synchronization, chaotic spiking, and~chaotic bursting while remaining small enough to allow efficient computation of full Lyapunov spectra and attractor dimensions. Importantly, we choose a number of neurons higher than, say, five neurons, because we require a high-dimensional state space to explore the relationship between the system's chaotic dynamics and the fractal geometry of its high-dimensional attractors. With~30 two-dimensional neurons (see Equation~\eqref{eq:ring-system-state-vector}), the~system evolves in a 60-dimensional state space, which is sufficient to host high-dimensional chaotic attractors whose fractal dimensions can span a wide range, revealing clearer trends. In~this paper, we explore three different regimes of the ring lattice~system:
\begin{enumerate}[leftmargin=*,labelsep=5mm]
    \item The homogeneous case, where all neurons have the same $\sigma_i$ and $\alpha_i$ values;
    \item The partially heterogeneous case, where each neuron has its own $\sigma_i$ value but the same $\alpha_i$ values;
    \item The fully heterogeneous case, where each neuron has its own $\sigma_i$ and $\alpha_i$ values.
\end{enumerate}

In all three cases, each neuron has a different initial $x$ value but the same initial $y$ value. We do not consider the case where each neuron has its own $y_{i,0}$ value because different evolutions of the slow variable are accounted for by different values of $\sigma_i$ \cite{rulkov}.

In Appendix \ref{appen:jacobian}, a~sketch of the derivation of the Jacobian matrix of the ring system is shown (Equation~\eqref{eq:THE-jacobian-entry}). Given some initial state $\mathbf{X}_0$, an~orbit $O(\mathbf{X}_0) = \{\mb{X}_0,\mb{X}_1,\hdots,\mb{X}_{999}\}$ of length 1000 (which is sufficiently long for Lyapunov exponent convergence) is generated, and the Jacobian matrix of the system $J(\mathbf{X})$ is calculated at each $\mathbf{X}\in O(\mathbf{X}_0)$. Then, the~QR factorization method detailed in Ref.~\cite{eckmann} and\mbox{ Appendix \ref{appen:qr}} for calculating the Lyapunov spectrum is used to compute the 60 Lyapunov exponents of the orbit. The~maximal Lyapunov exponent is used to gauge chaotic dynamics in this section, and~the entire Lyapunov spectrum is used for the analysis in Section~\ref{sec:geometry}. Specifically, in~this paper, we adopt the definition of chaos from Ref.~\cite{alligood}, which characterizes a system as chaotic if its maximal Lyapunov exponent is greater than~zero. 

We will now present our results detailing the dynamics that emerge from the homogeneous regime of the ring system. We choose the parameters $\sigma_i = -0.5$ and $\alpha_i = 4.5$ for all of the neurons, which set the individual neurons in the nonchaotic spiking regime. Additionally, the~initial slow-variable values for all of the neurons are set to $y_{i,0} = -3.25$. However, setting the initial fast-variable values to be equal would be pointless because the neurons would have identical dynamics, resulting in no current flow between them. Instead, $x_{i,0}$ variables are randomly chosen from the interval $(-1,1)$. The specific random initial states and parameters used are listed in Appendix \ref{appen:supp}.

In Figure~\ref{fig:random_x_graphs}, the~first thousand iterations of the fast-variable orbits of the first eight Rulkov neurons in the ring are graphed. We start with uncoupled neurons $g=0$ in Figure~\ref{fig:random_x_graphs}a, where uncoupled neurons with identical parameters are all out of phase in the nonchaotic spiking domain. As~expected, because~there is no current flow and all of the individual Rulkov neurons are spiking regularly, the~maximal Lyapunov exponent $\lambda_1$ is negative. When the electrical coupling strength is raised to $g=0.05$ (Figure~\ref{fig:random_x_graphs}b), the~neurons still spike relatively periodically, but~there are some irregularities when one voltage happens to catch onto another. This small $g$ is enough to make the system chaotic, with~$\lambda_1\approx 0.0491>0$. Next, the~coupling strength is raised significantly to $g=0.25$, where the ring system now exhibits synchronized chaotic bursting (Figure~\ref{fig:random_x_graphs}c). Here, synchronization refers to the oscillations between rapid spiking and silence happening in sync with each other. However, aligning with other computational neuron modelings, the~individual spikes within the bursts are chaotic and unsynchronized~\cite{rulkov, rulkov2}. Finally, the~coupling strength is taken to the extreme with $g=1$ in Figure~\ref{fig:random_x_graphs}d, where synchronized hyperchaos ensues ($\lambda_1\approx 0.1694$) due to each Rulkov neuron having an overwhelming influence on its nearest neighbors. The~use of the term ``hyperchaos,'' generally defined to be chaotic dynamics with at least two positive Lyapunov exponents~\cite{hyperchaos}, is justified here because the orbit in Figure~\ref{fig:random_x_graphs}d has 11 positive Lyapunov exponents (out of 60).

\begin{figure}[H]
    
    \includegraphics[scale=0.7]{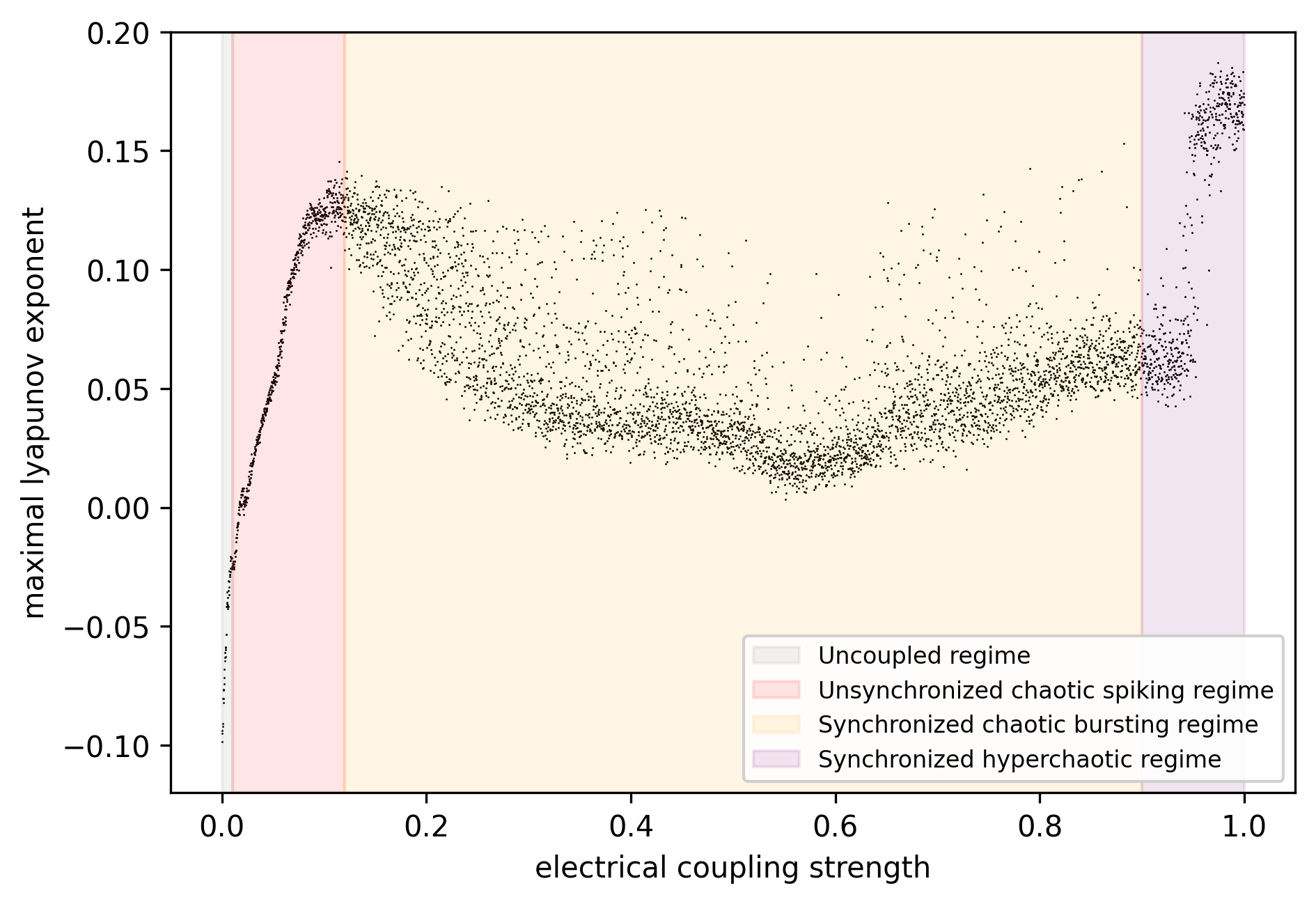}
    \caption{Graph of the maximal Lyapunov exponent $\lambda_1$ against the electrical coupling strength $g$ for the homogeneous case, with~$x_{i,0}\in (-1,1)$, $y_{i,0}=-3.25$, $\sigma_i = -0.5$, and~$\alpha_i = 4.5$. The~maximal Lyapunov exponent graph shows the four distinct regimes of behavior: the uncoupled regime, unsynchronized chaotic spiking regime, synchronized chaotic bursting regime, and~synchronized hyperchaotic regime. The~maximal Lyapunov exponents $\lambda_1$ are calculated using orbits of length 1000, which is sufficient for~convergence.}
    \label{fig:max_lyap_exp_graph_random_x}
\end{figure}

A natural question to ask is how the maximal Lyapunov exponent changes as $g$ is varied, a~graph of which is displayed in Figure~\ref{fig:max_lyap_exp_graph_random_x} for this homogeneous case. In~the figure, 5000 evenly spaced values of $g$ between 0 and 1 are considered, and~each point represents the maximal Lyapunov exponent of the system for one of these $g$ values. We notice that the maximal Lyapunov exponents are rather erratic for $g>0.1$, covering a wide range of values over a small domain of $g$ values. However, there do exist some general trends. Because~the individual neurons in this system are nonchaotic, $\lambda_1$ values initially start below zero. As~the current starts to flow, the~range of chaotic spiking is reached (e.g., Figure~\ref{fig:random_x_graphs}b), where the $\lambda_1$ values quickly become positive and reach a maximum. Then, as~the synchronized chaotic bursting regime is reached (e.g., Figure~\ref{fig:random_x_graphs}c), the~$\lambda_1$ values become much more erratic but exhibit an overall downward trend, which can be attributed to the nonchaotic silence between bursts of spikes. As~the extreme values of $g$ towards the right side of the graph are reached (e.g., Figure~\ref{fig:random_x_graphs}d), $\lambda_1$ shoots up to high and hyperchaotic~values.

\begin{figure}[H]
    \begin{adjustwidth}{-\extralength}{0cm}
    \centering
    \subfloat[\centering]{\includegraphics[scale=0.63]{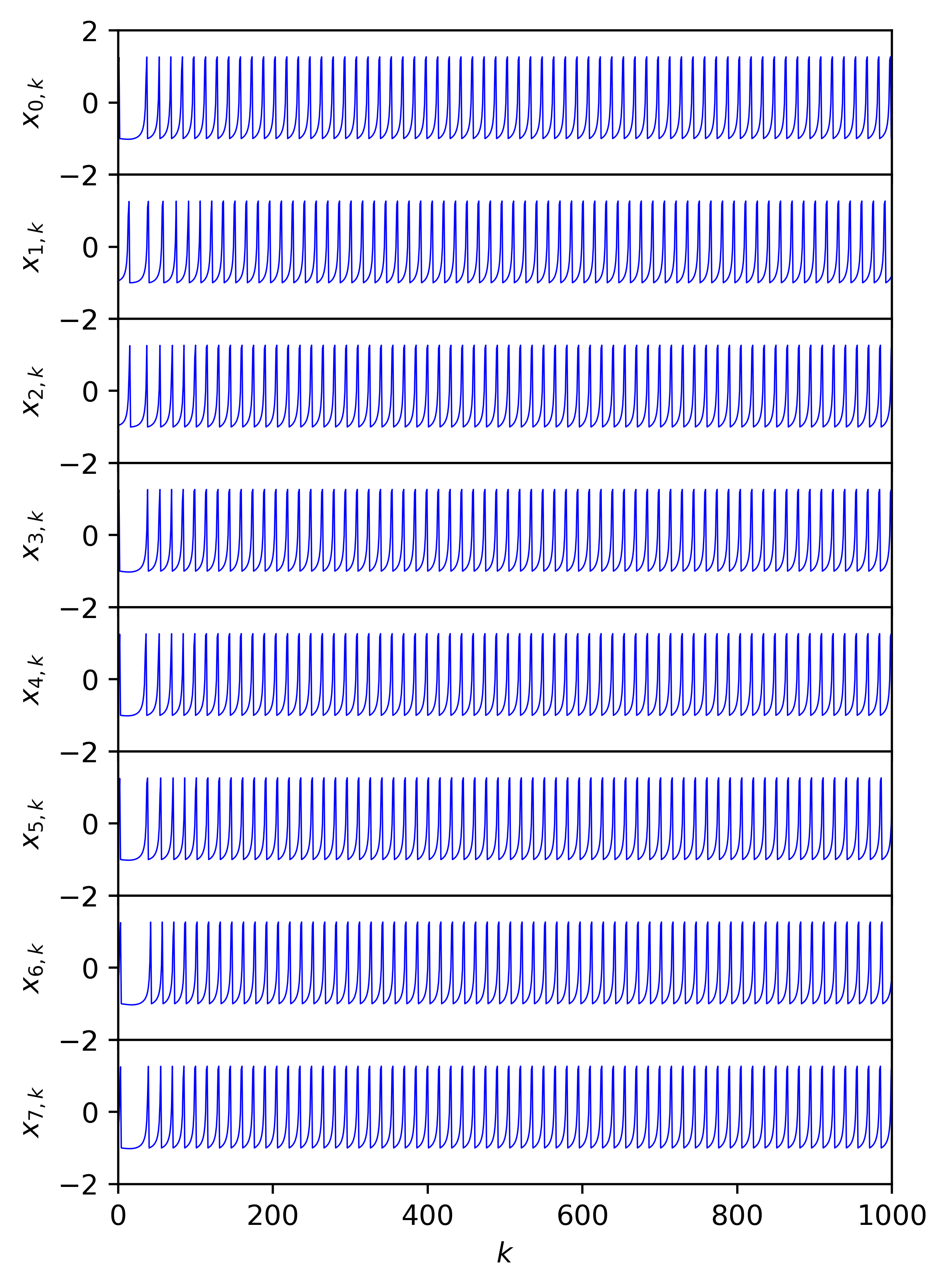} }
    \subfloat[\centering]{\includegraphics[scale=0.63]{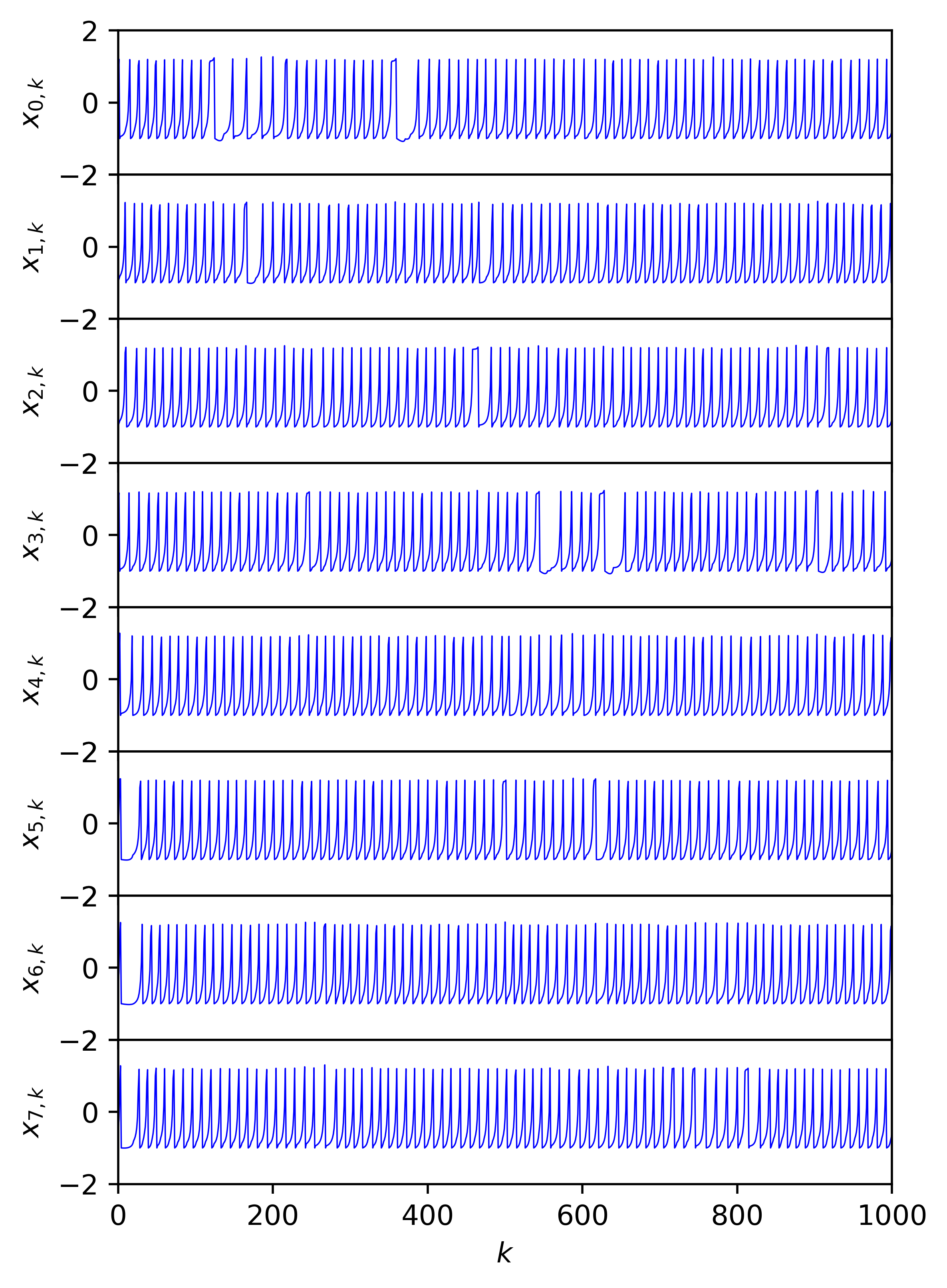}} \\
    \subfloat[\centering]{\includegraphics[scale=0.63]{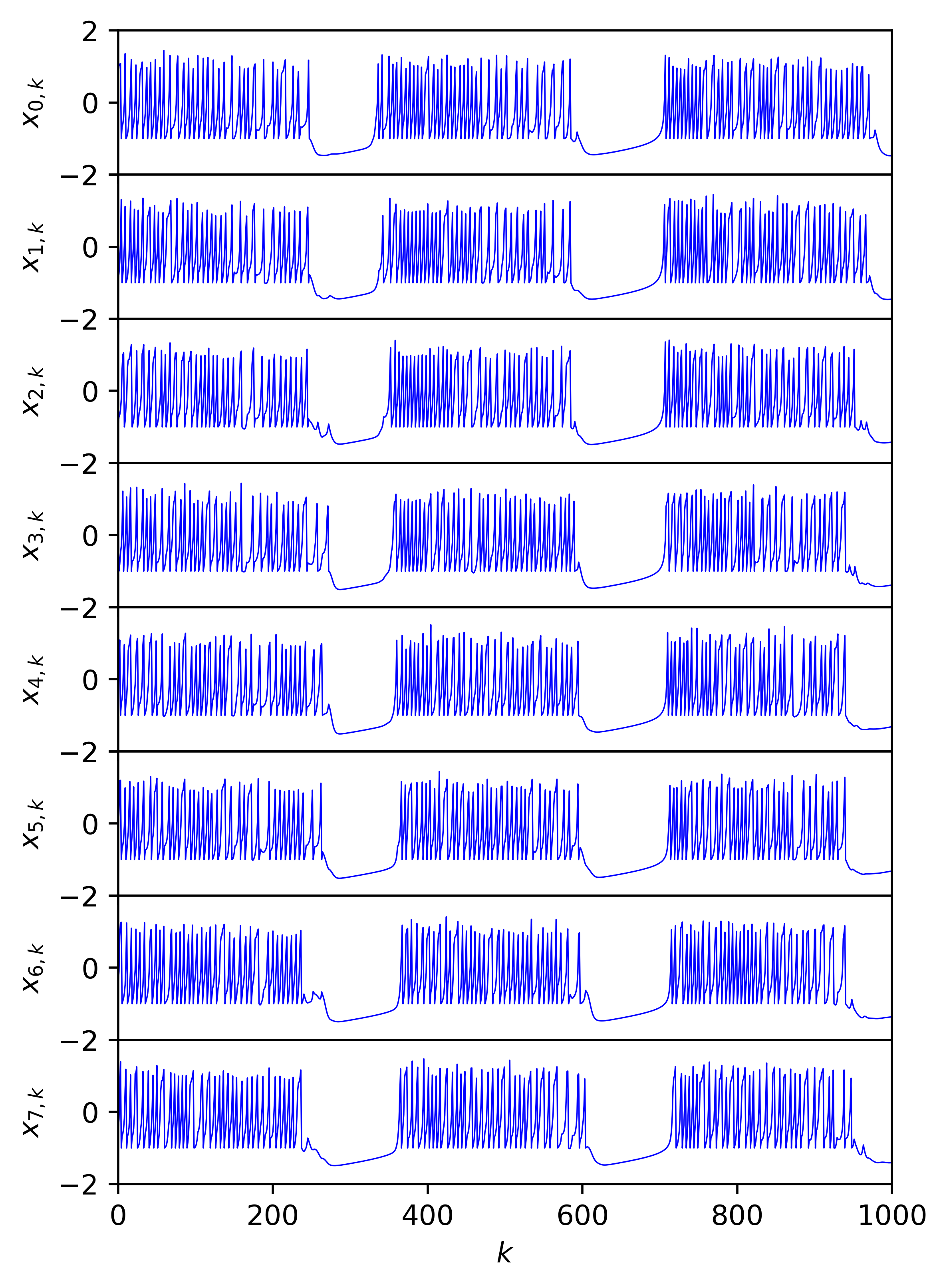} }
    \subfloat[\centering]{\includegraphics[scale=0.63]{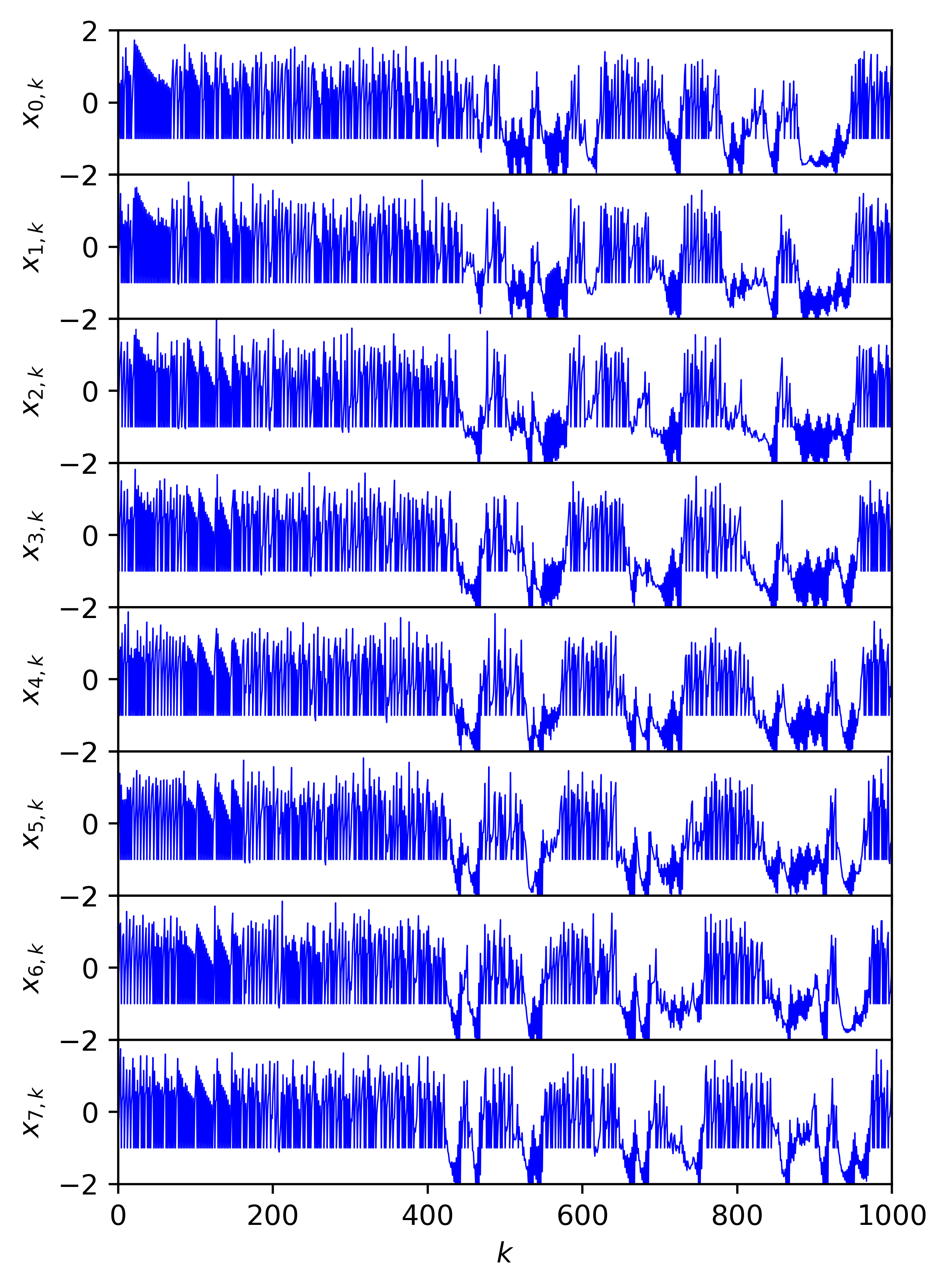} }
    \end{adjustwidth}
    \caption{Graphs of the fast-variable orbits of the first eight neurons in the homogeneous regime of the ring lattice system, with~$x_{i,0}\in (-1,1)$, $y_{i,0}=-3.25$, $\sigma_i = -0.5$, and~$\alpha_i = 4.5$. The~four coupling strength values show four distinct regimes of behavior:
    (\textbf{a}) $g=0$, $\lambda_1\approx -0.0938$ (uncoupled nonchaotic spiking); (\textbf{b}) $g=0.05$, $\lambda_1\approx 0.0491$ (unsynchronized chaotic spiking); (\textbf{c}) $g=0.25$, $\lambda_1\approx 0.0595$ (synchronized chaotic bursting); and (\textbf{d}) $g=1$, $\lambda_1\approx 0.1694$ (synchronized hyperchaos).}
    \label{fig:random_x_graphs}
\end{figure}
\unskip

\begin{figure}[H]
    \begin{adjustwidth}{-\extralength}{0cm}
    \centering
    \subfloat[\centering]{\includegraphics[scale=0.43]{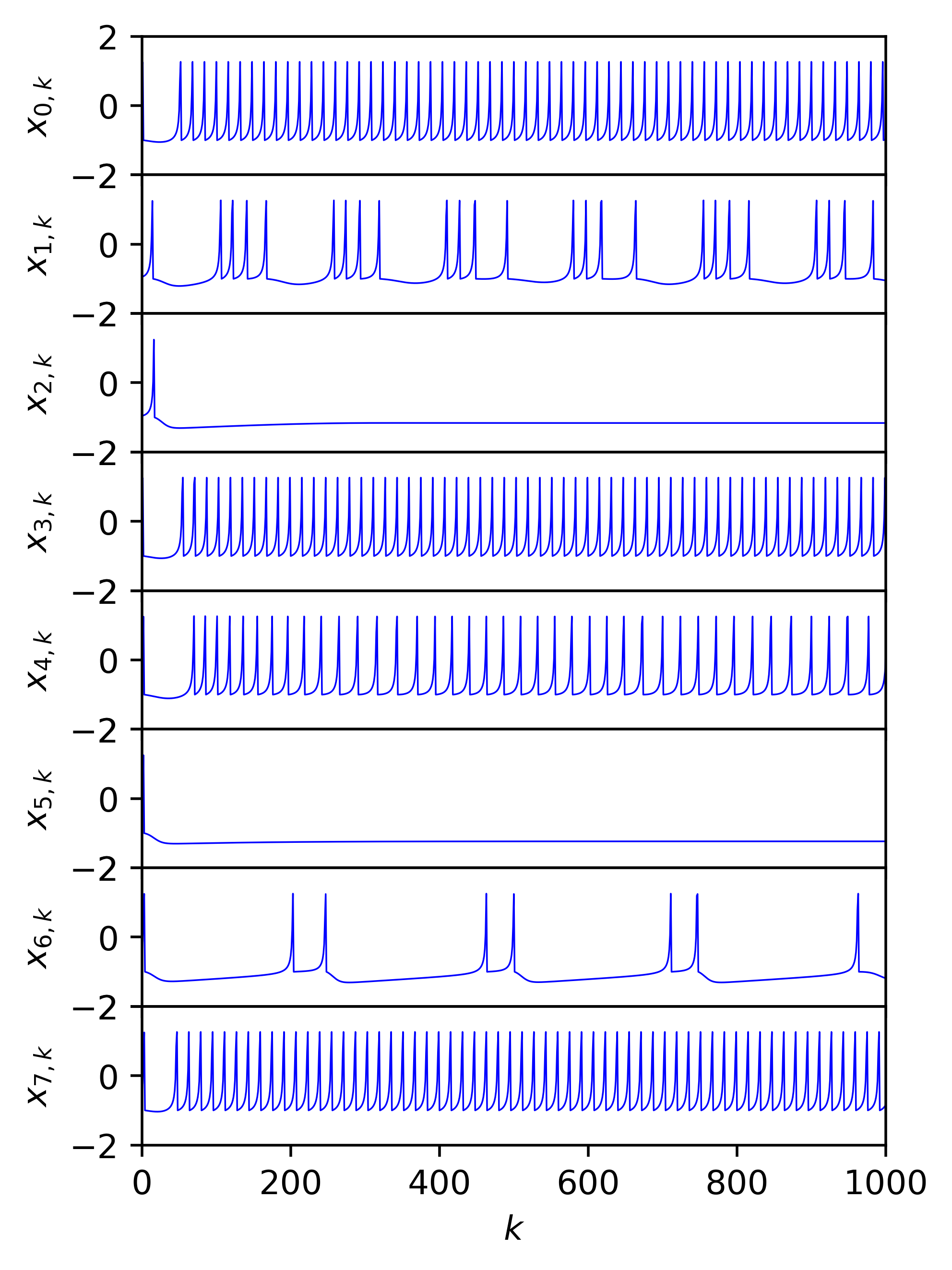} \label{fig:random_sigma_ge0}}
    \subfloat[\centering]{\includegraphics[scale=0.43]{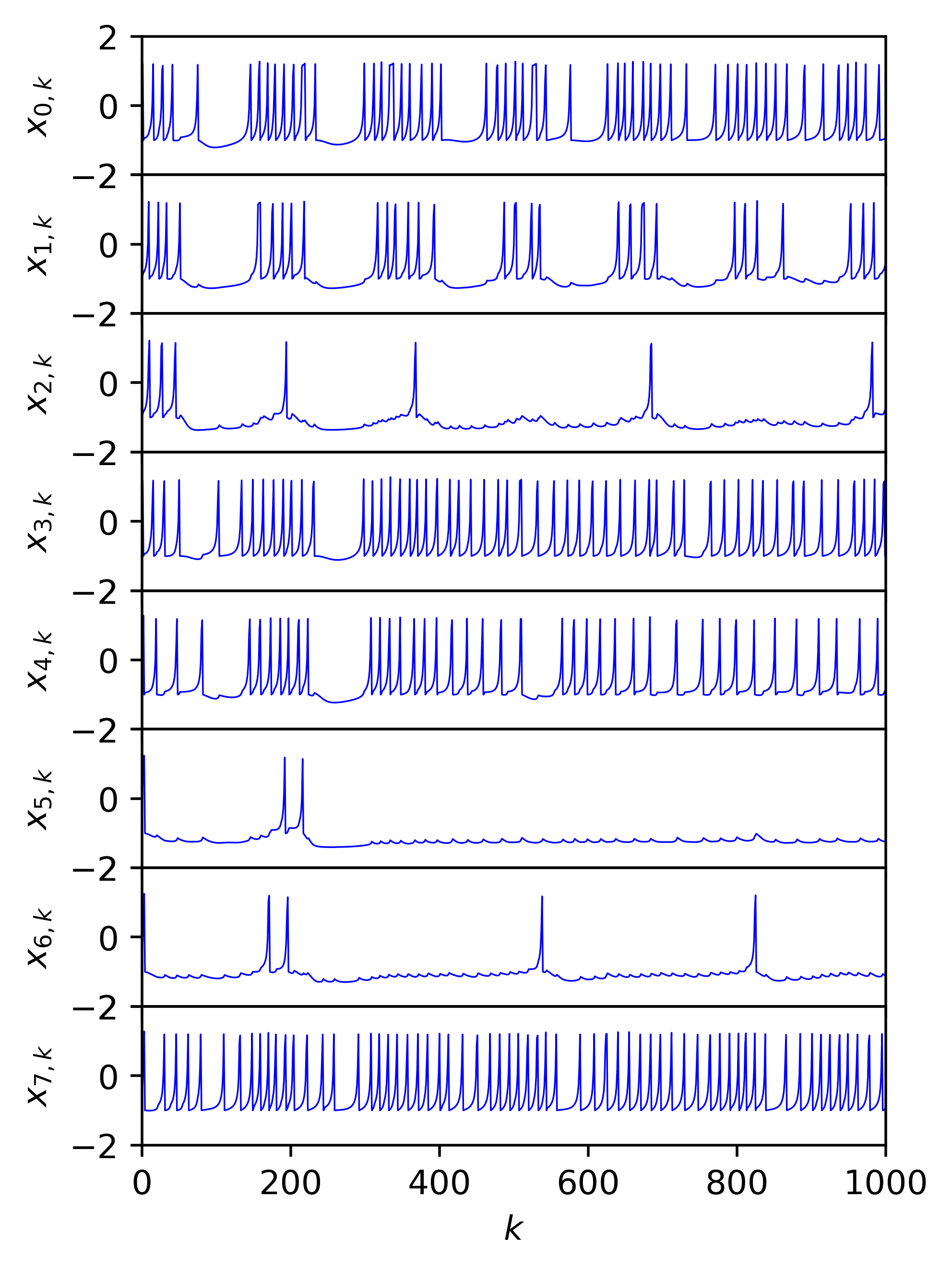} \label{fig:random_sigma_ge0.05}}
    \subfloat[\centering]{\includegraphics[scale=0.43]{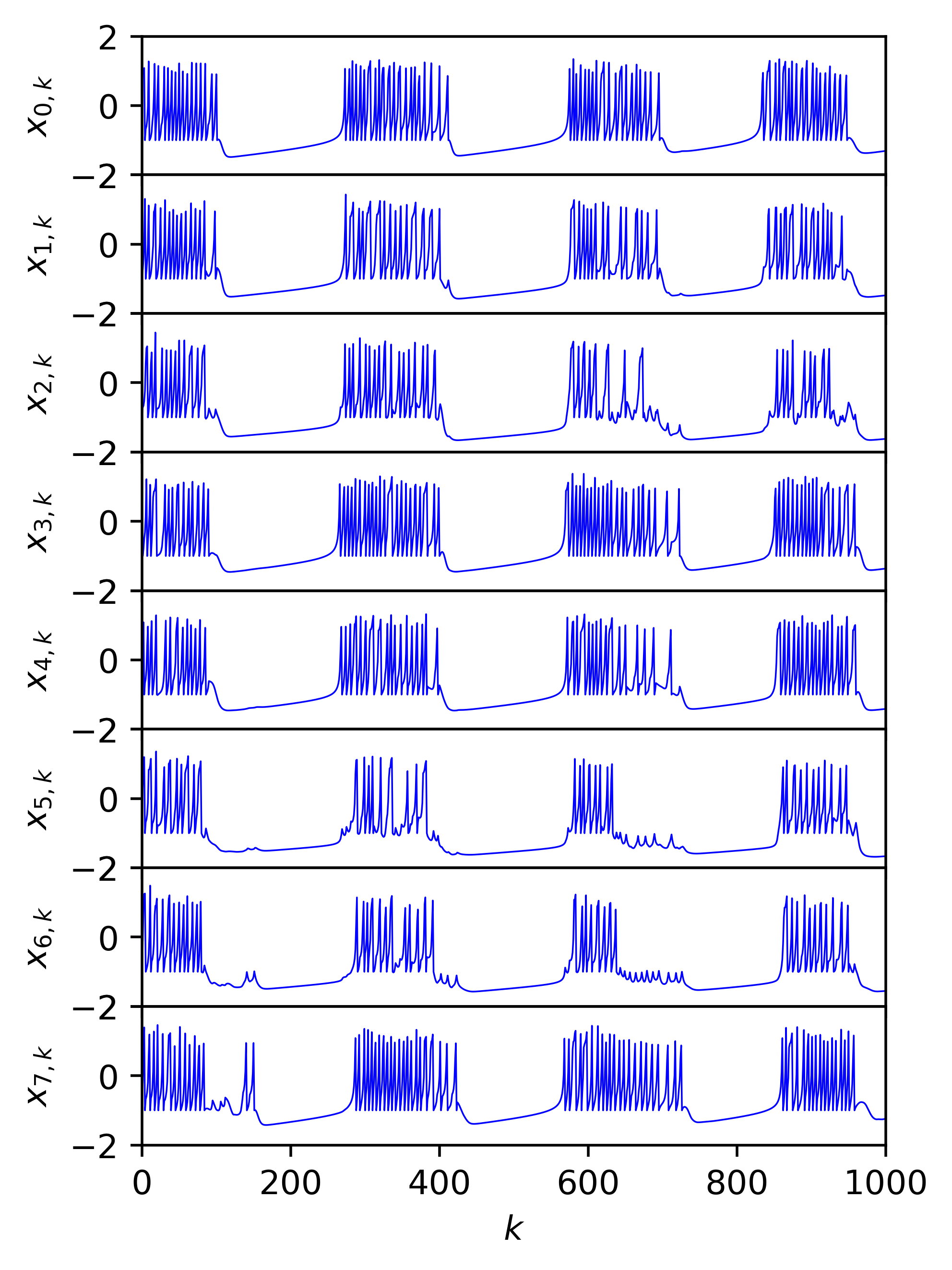} \label{fig:random_sigma_ge0.25}}
    \subfloat[\centering]{\includegraphics[scale=0.43]{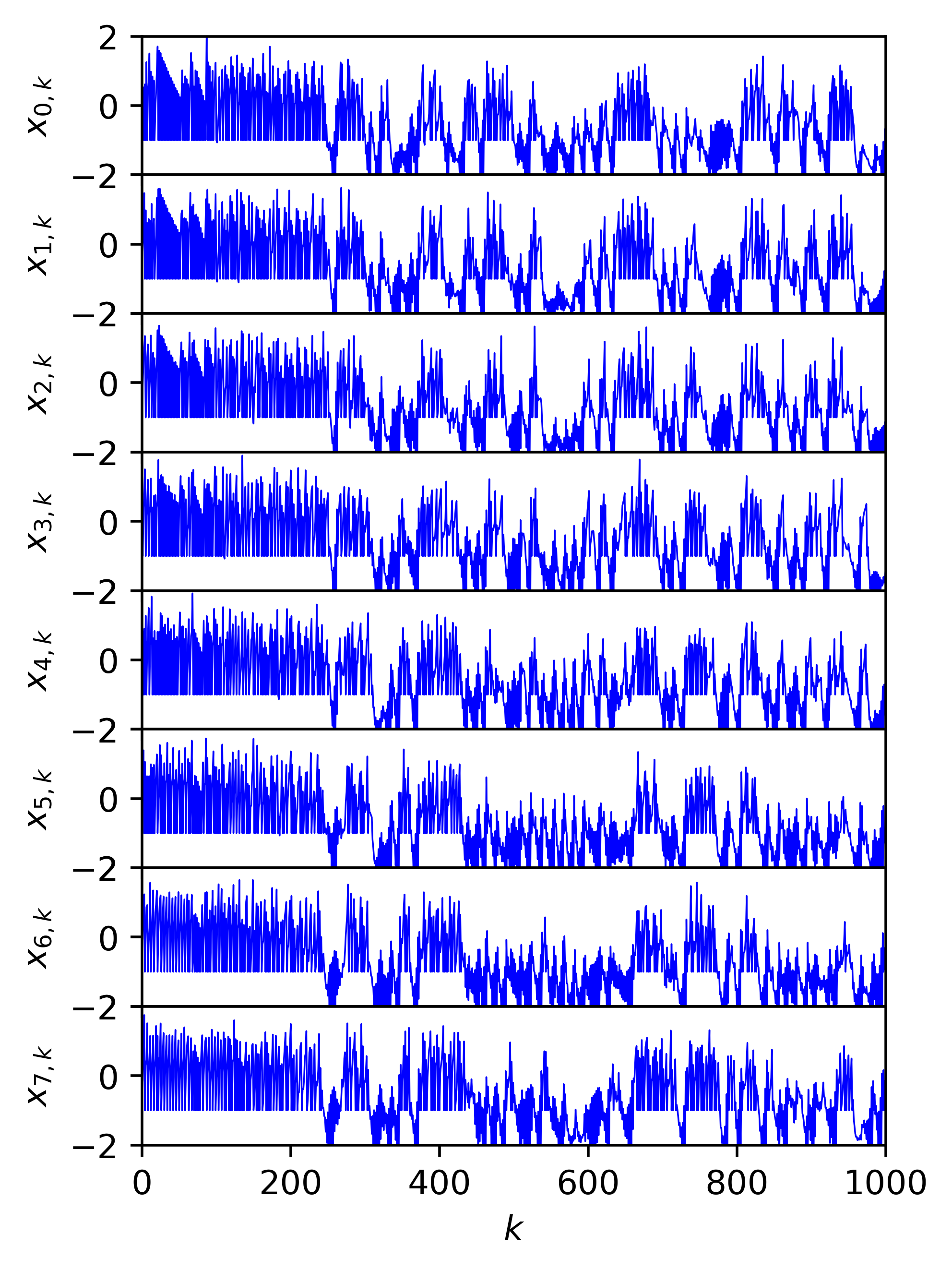} \label{fig:random_sigma_ge1}}
    \end{adjustwidth}
    \caption{Graphs of the fast-variable orbits of the first eight neurons in the partially heterogeneous regime of the ring lattice system, with~$x_{i,0}\in (-1,1)$, $y_{i,0}=-3.25$, $\sigma_i\in(-1.5,-0.5)$,\linebreak   and~$\alpha_i = 4.5$. The~four coupling strength values show four distinct regimes of behavior: (\textbf{a}) $g=0$, $\lambda_1\approx 0.0644$ (uncoupled regime); (\textbf{b}) $g=0.05$, $\lambda_1\approx 0.0686$ (weakly coupled regime); (\textbf{c}) $g=0.25$, $\lambda_1\approx 0.0663$ (synchronized chaotic bursting regime); and (\textbf{d}) $g=1$, $\lambda_1\approx 0.2003$ (synchronized \mbox{hyperchaotic regime).}}
    \label{fig:random_sigma_graphs}
\end{figure}

The partially and fully heterogeneous cases, in~which different neurons in the ring have different parameters, will now be examined. The~partially heterogeneous case keeps the same randomly distributed $x_{i,0}$ values (Equation~\eqref{eq:big-initial-state}), the~same $y_{i,0}=-3.25$ values, and~the same $\alpha_i = 4.5$ values, but~it has randomly chosen $\sigma_i$ values from the interval $(-1.5,-0.5)$ (Equation~\eqref{eq:big-sigma-vector}). With~these parameters, different individual neurons are in the silence, spiking, and~bursting domains~\cite{rulkov}, which can be seen in the visualization of the uncoupled neuron system's dynamics (Figure~\ref{fig:random_sigma_graphs}a). 
\vspace
{-6pt}

\begin{figure}[H]
    \begin{adjustwidth}{-\extralength}{0cm}
    \centering
    \subfloat[\centering]{\includegraphics[scale=0.43]{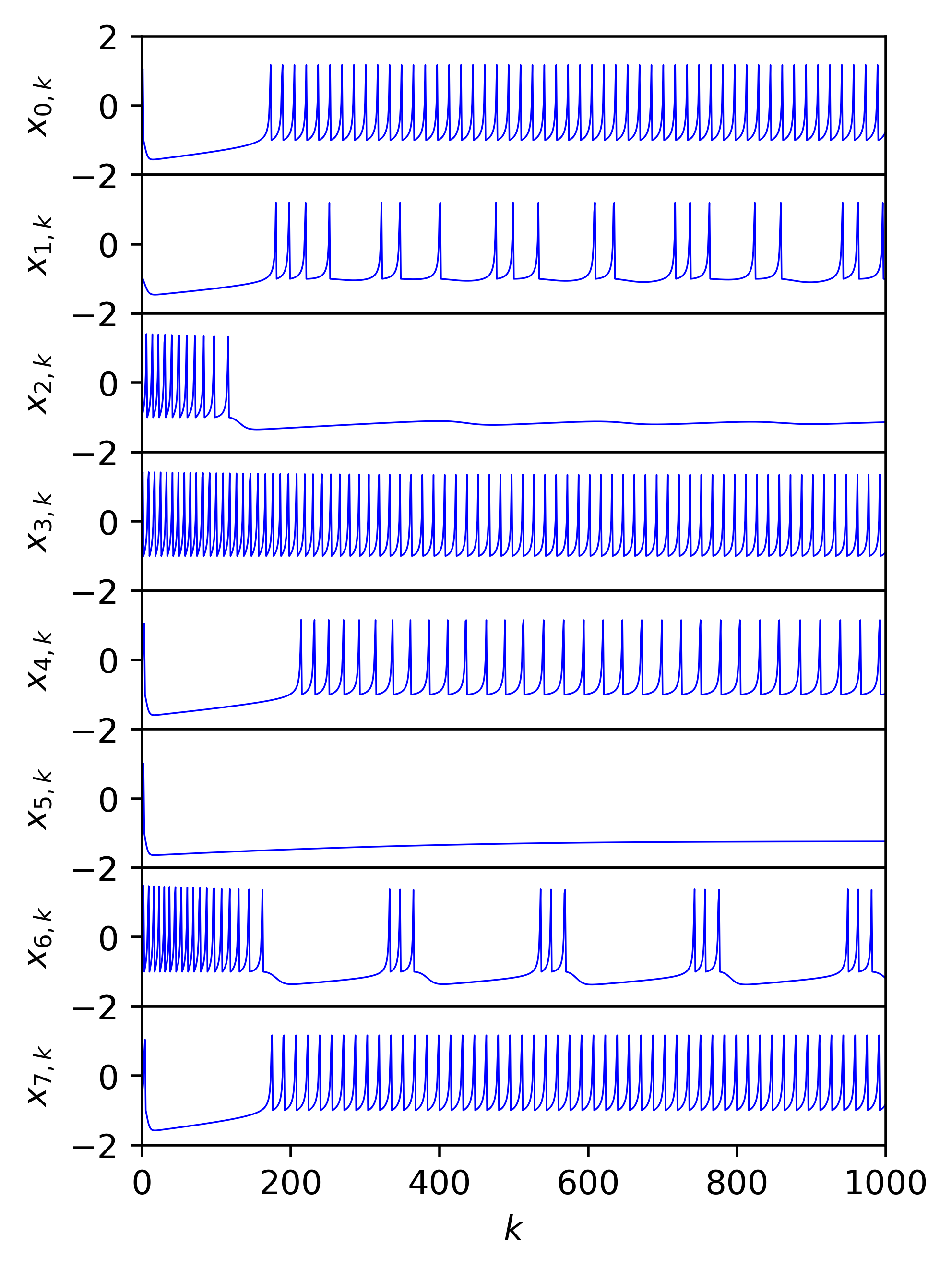} \label{fig:random_alpha_ge0}}
    \subfloat[\centering]{\includegraphics[scale=0.43]{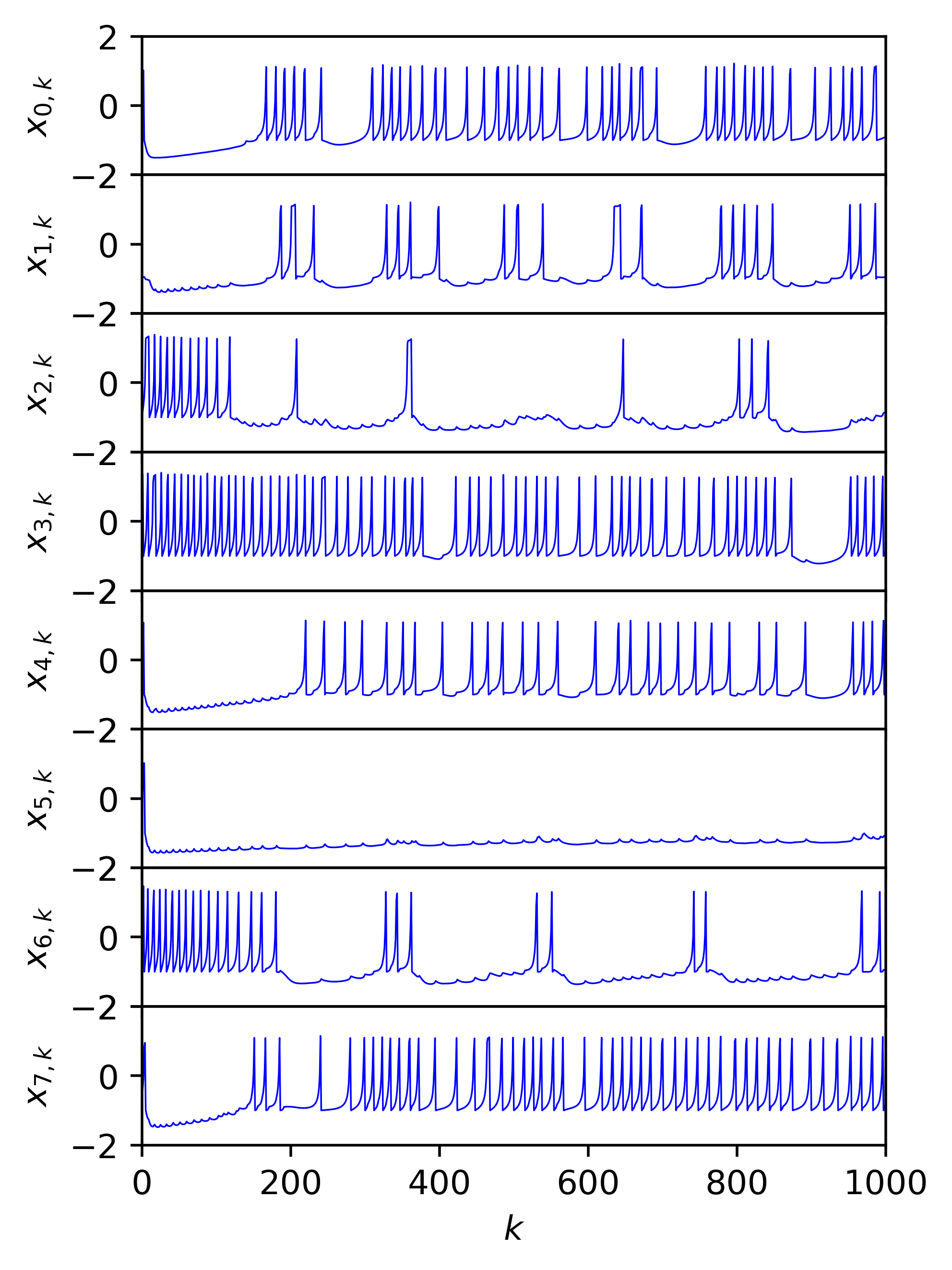} \label{fig:random_alpha_ge0.05}}
    \subfloat[\centering]{\includegraphics[scale=0.43]{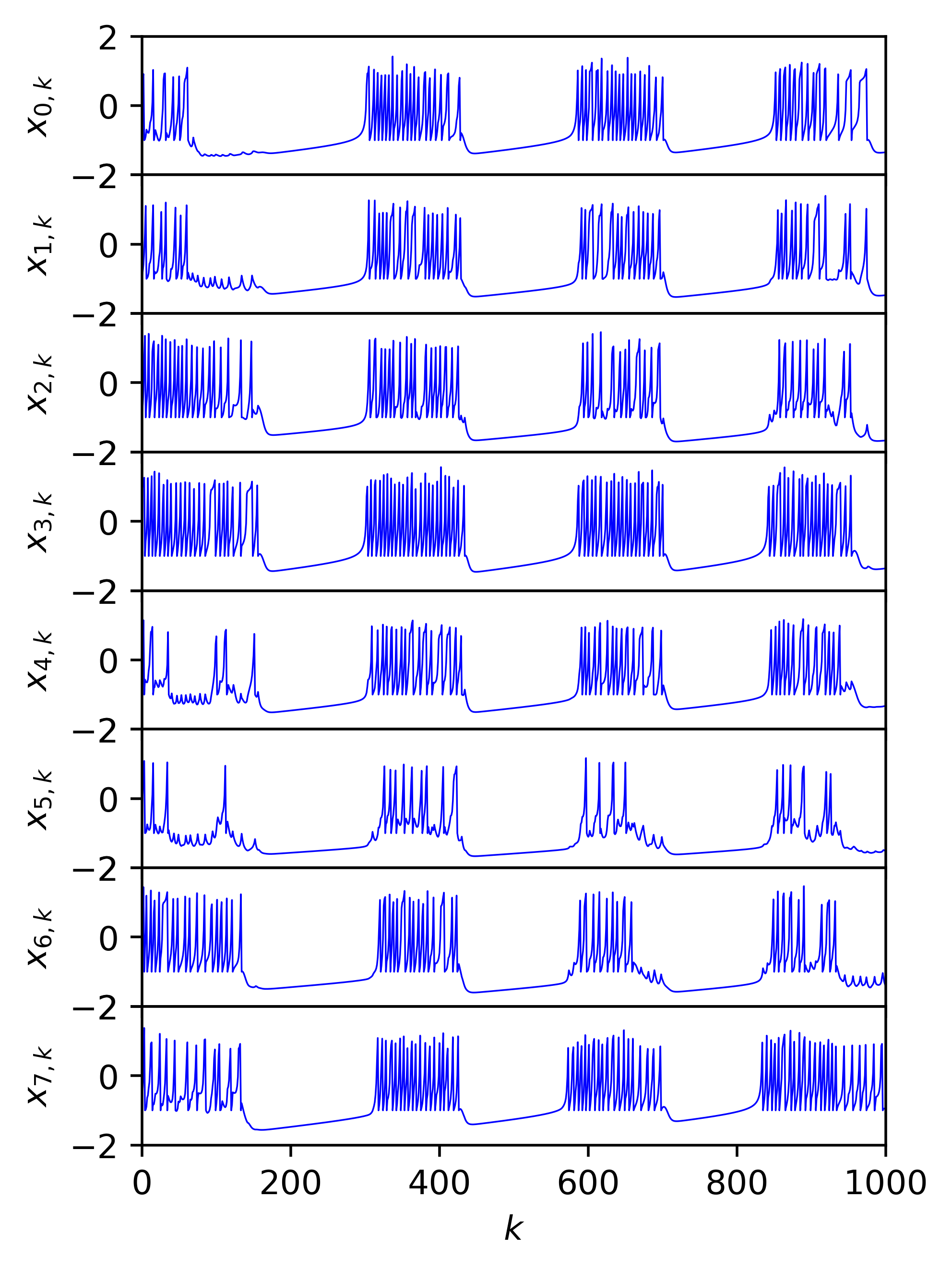} \label{fig:random_alpha_ge0.25}}
    \subfloat[\centering]{\includegraphics[scale=0.43]{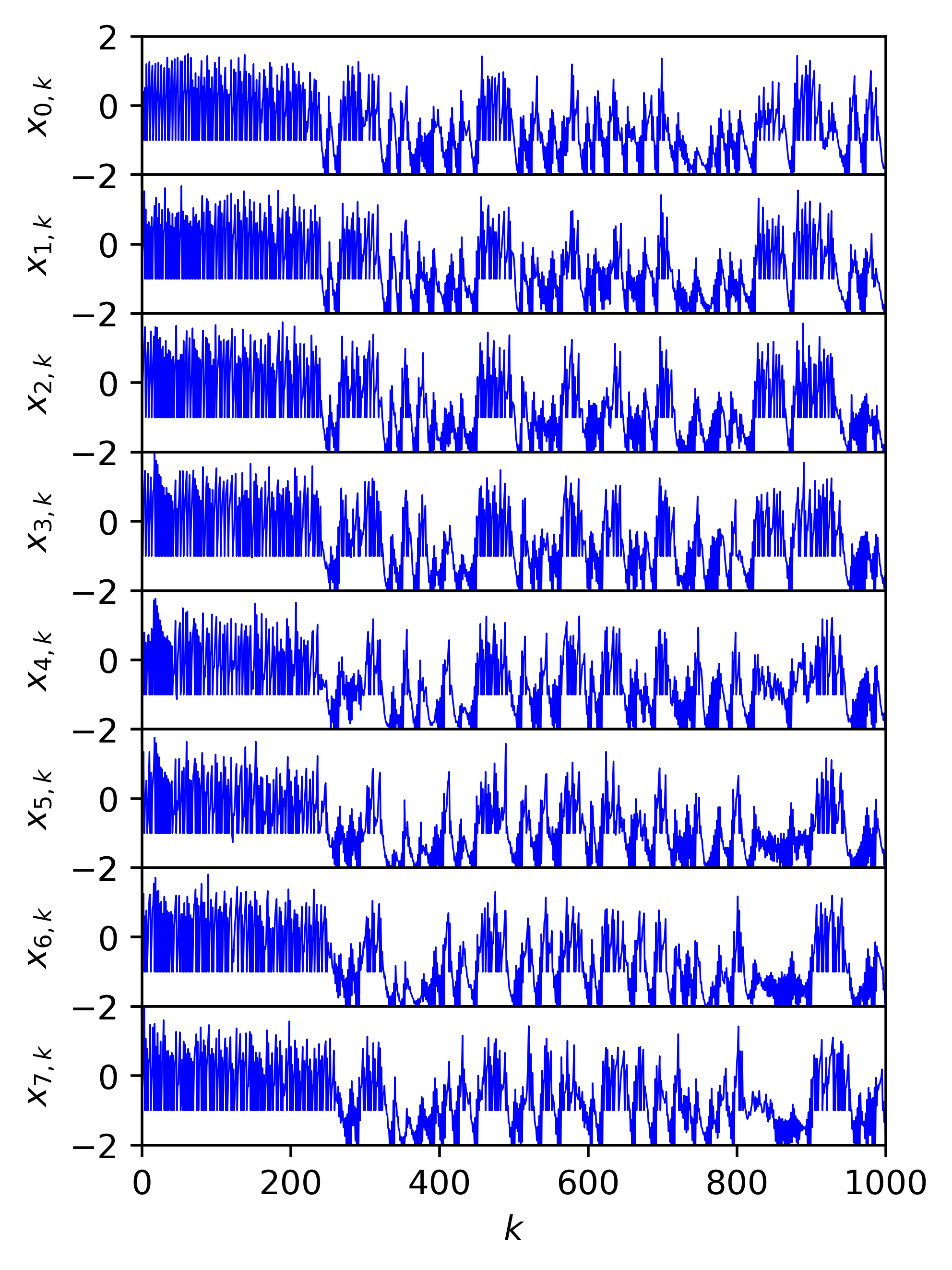} \label{fig:random_alpha_ge1}}
    \end{adjustwidth}
    \caption{Graphs of the fast-variable orbits of the first eight neurons in the fully heterogeneous regime of the ring lattice system, with~$x_{i,0}\in (-1,1)$, $y_{i,0}=-3.25$, $\sigma_i\in(-1.5,-0.5)$, and~$\alpha_i\in (4.25,4.75)$. The~four coupling strength values show four distinct regimes of behavior: (\textbf{a}) $g=0$, $\lambda_1\approx 0.0469$ (uncoupled regime); (\textbf{b}) $g=0.05$, $\lambda_1\approx 0.0563$ (weakly coupled regime); (\textbf{c}) $g=0.25$, $\lambda_1\approx 0.0633$ (synchronized chaotic bursting regime); (\textbf{d}) $g=1$, $\lambda_1\approx 0.2053$ (synchronized \mbox{hyperchaotic regime).}}
    \label{fig:random_alpha_graphs}
\end{figure}

Finally, the~third regime we analyze is the fully heterogeneous case, where we keep the randomly distributed $x_{i,0}$ and $\sigma_i$ values and keep $y_{i,0}=-3.25$, but choose random $\alpha_i$ values from the interval $(4.25,4.75)$ (Equation~\eqref{eq:big-alpha-vector}). This further varies the distribution of possible behaviors between different neurons in the system. This can be seen in the dynamics of the uncoupled neuron system (Figure~\ref{fig:random_alpha_graphs}a), where some neurons exhibit rapid spiking, some burst occasionally, and~some are~silent. 
\vspace{-8pt}

In Figures~\ref{fig:random_sigma_graphs} and \ref{fig:random_alpha_graphs}, the~fast-variable orbits of the first eight neurons in the ring are graphed using the same electrical coupling strength values as the homogeneous case: $g=0,0.05,0.25,1$. Comparing both of these regimes to the homogeneous case, similar patterns emerge among them. For~$g=0.05$, the~adjacent neurons start to have some effect on each other, but~the overall dynamical picture remains the same. Upon raising the electrical coupling strength up to $g=0.25$, all the neurons undergo synchronized chaotic bursting, and~upon going to the extreme $g=1$, synchronized hyperchaos ensues. An~interesting observation that is even clearer in these visualizations is neurons' direct influence on their adjacent partners. For~instance, in~Figures~\ref{fig:random_sigma_graphs}b and \ref{fig:random_alpha_graphs}c, spiking in one neuron is reflected in adjacent neurons with smaller spikes during a period of~silence.

\begin{figure}[H]
    \begin{adjustwidth}{-\extralength}{0cm}
    \centering
    \subfloat[\centering]{\includegraphics[scale=0.575]{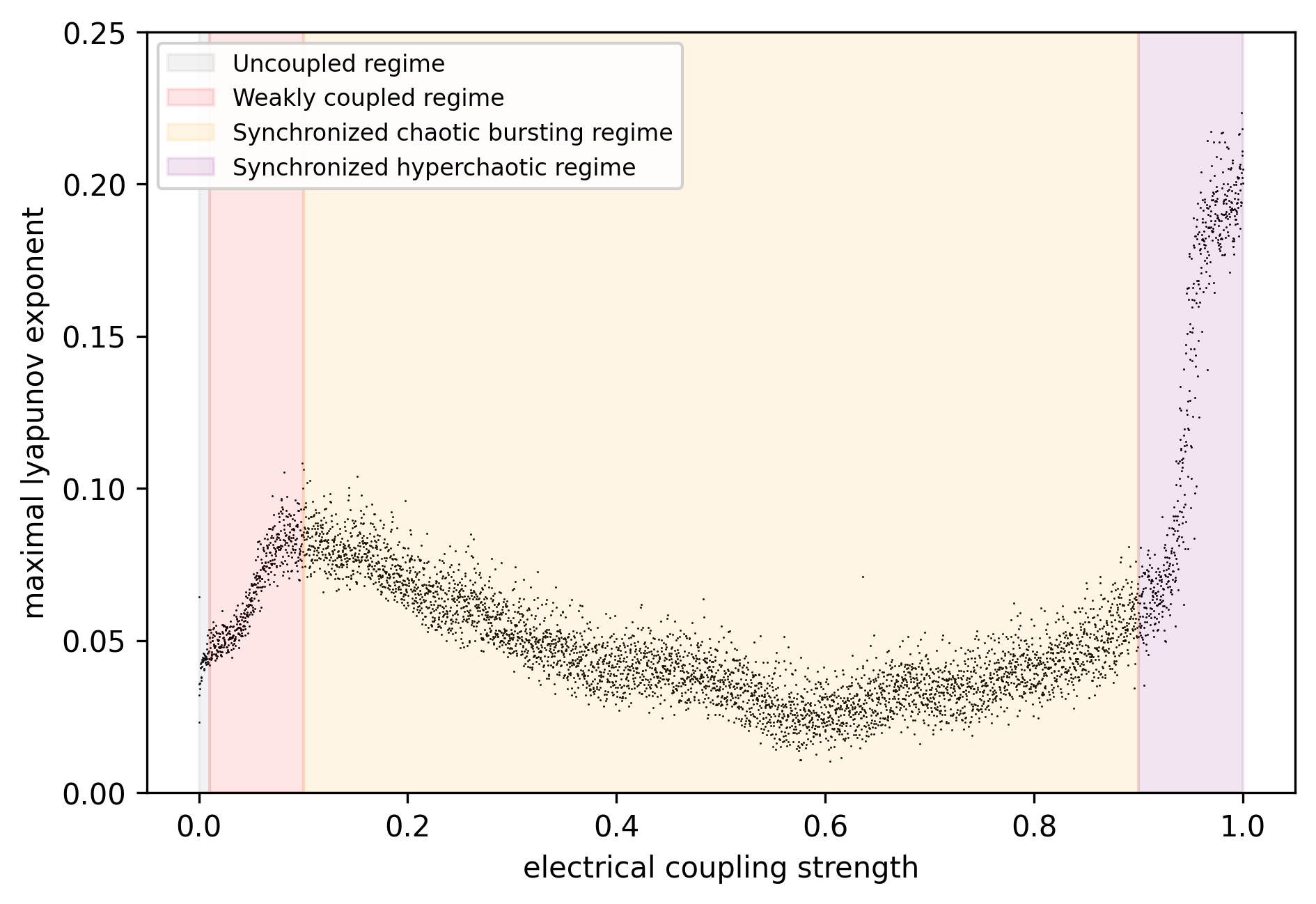} \label{fig:ring-max-lyap-random-sigma}}
    \subfloat[\centering]{\includegraphics[scale=0.575]{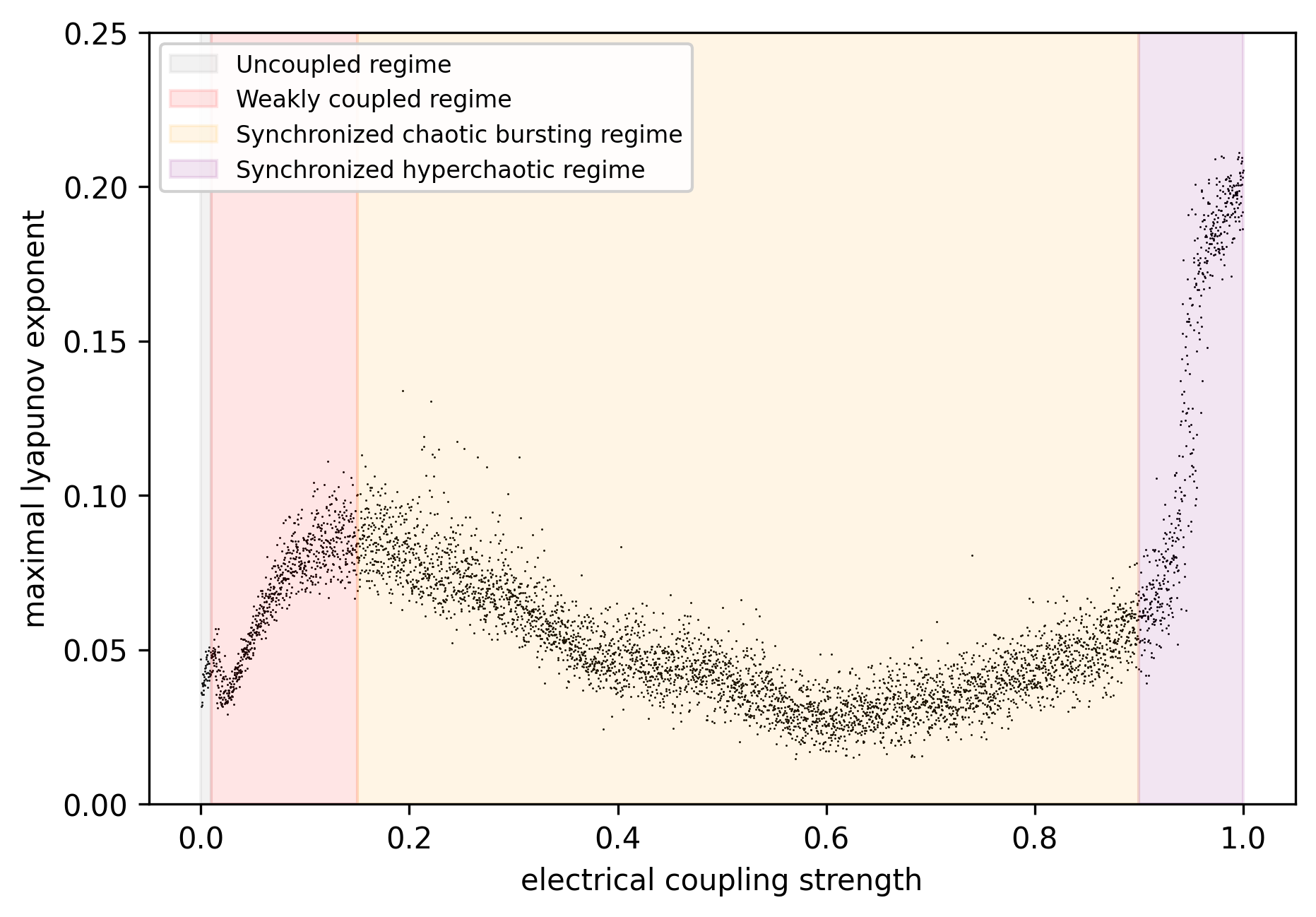} \label{fig:ring-max-lyap-random-alpha}}
    \end{adjustwidth}
    \caption{Graphs of the maximal Lyapunov exponent $\lambda_1$ against the electrical coupling strength $g$ for (\textbf{a}) the partially heterogeneous case, with~$x_{i,0}\in (-1,1)$, $y_{i,0}=-3.25$, $\sigma_i\in(-1.5,-0.5)$, and~$\alpha_i = 4.5$, and~(\textbf{b}) the fully heterogeneous case, with~$x_{i,0}\in (-1,1)$, $y_{i,0}=-3.25$, $\sigma_i\in(-1.5,-0.5)$, and~$\alpha_i\in (4.25,4.75)$. The~maximal Lyapunov exponent graphs for the two cases are similar, showing the same four distinct regimes of behavior: the uncoupled regime, weakly coupled regime, synchronized chaotic bursting regime, and~synchronized hyperchaotic regime. The~maximal Lyapunov exponents $\lambda_1$ are calculated using orbits of length 1000, which is sufficient for~convergence.}
    \label{fig:ring-max-lyap-graphs-rand-sigandalph}
\end{figure}

Figure~\ref{fig:ring-max-lyap-graphs-rand-sigandalph} presents a visualization of the maximal Lyapunov exponents of these two regimes for many values of $g$. An~evident difference when comparing these graphs to the graph in Figure~\ref{fig:max_lyap_exp_graph_random_x} is that $\lambda_1>0$ for all the $g$ values. This is because even when the neurons are uncoupled, some of the individual neurons in the ring are chaotic. However, the~graphs of the maximal Lyapunov exponents for all three cases have similar shapes, the~major differences being when the neurons are weakly coupled and operating under their own parameters. Past this weak coupling domain, all three graphs in Figures~\ref{fig:max_lyap_exp_graph_random_x} and \ref{fig:ring-max-lyap-graphs-rand-sigandalph} follow the same increase up to chaotic spiking, followed by a swoop down as synchronized chaotic bursts occur, followed by a sharp increase as the extreme values of $g$ are approached. 
Therefore, despite making individual neurons exhibit drastically different dynamics from their neighbors, coupling makes the system exhibit similar dynamics. Although~this behavior has been observed to a lesser extent before in a Rulkov neuron system (see Ref.~\cite{osipov}), these distributions of Lyapunov exponents provide quantitative support for this~phenomenon.
\vspace{-6pt}


\section{Fractal Geometry of~Attractors}
\label{sec:geometry}

\begin{figure}[H]
    \begin{adjustwidth}{-\extralength}{0cm}
    \centering
    \subfloat[\centering]{\includegraphics[scale=0.55]{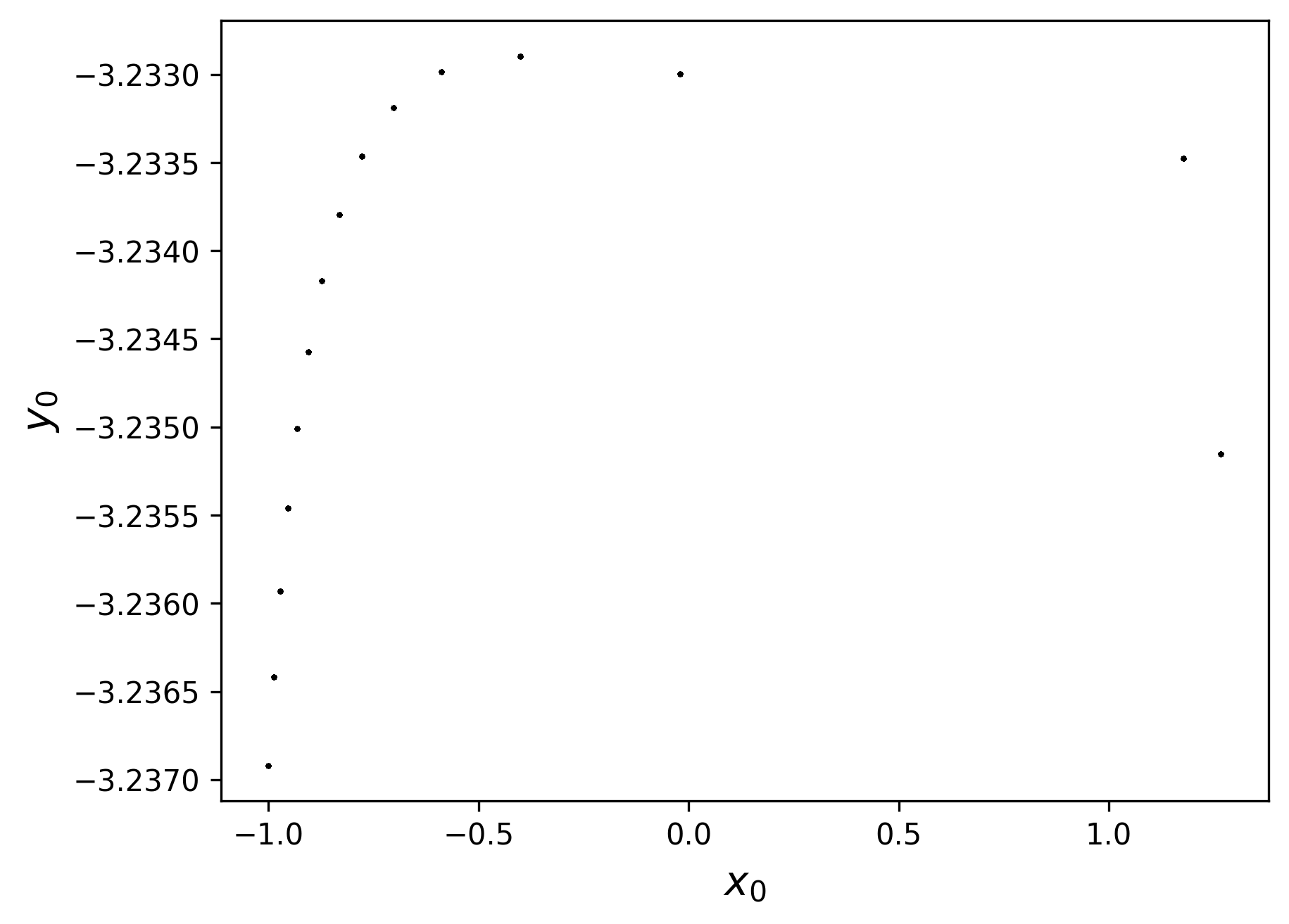} \label{fig:ring-slice-g0}}
    \subfloat[\centering]{\includegraphics[scale=0.55]{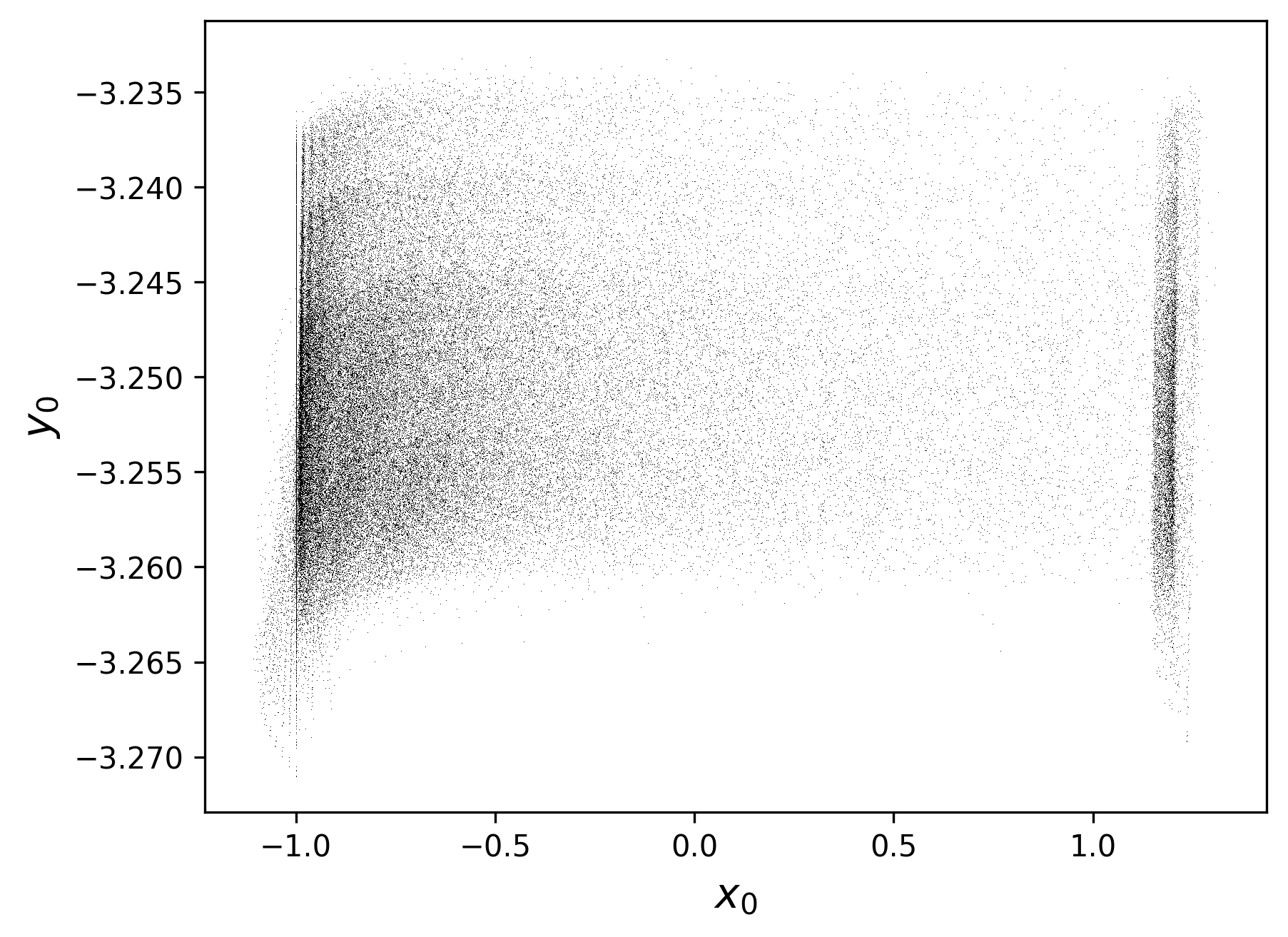} \label{fig:ring-slice-g0.05}} \\
    \subfloat[\centering]{\includegraphics[scale=0.55]{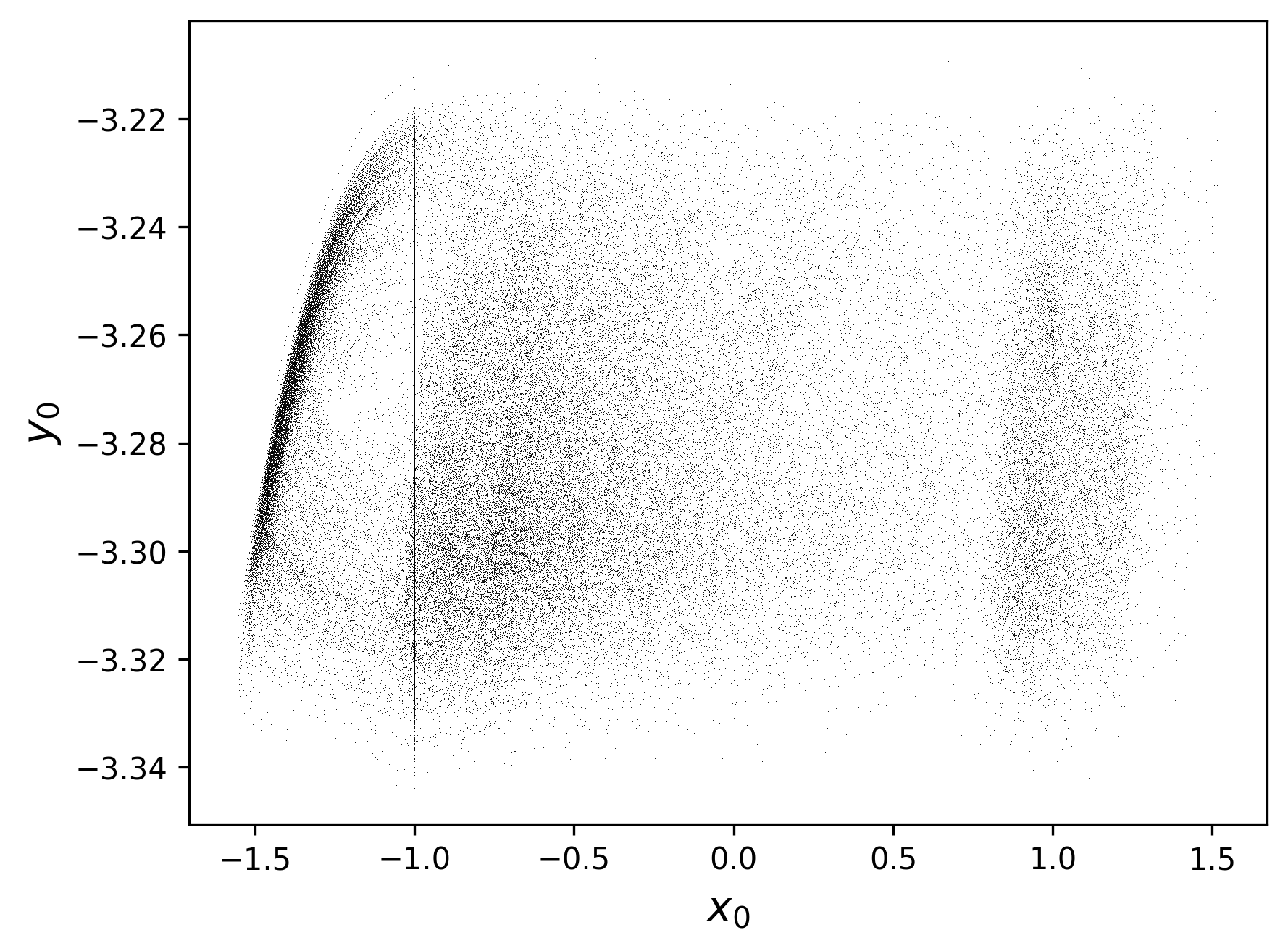} \label{fig:ring-slice-g0.25}}
    \subfloat[\centering]{\includegraphics[scale=0.55]{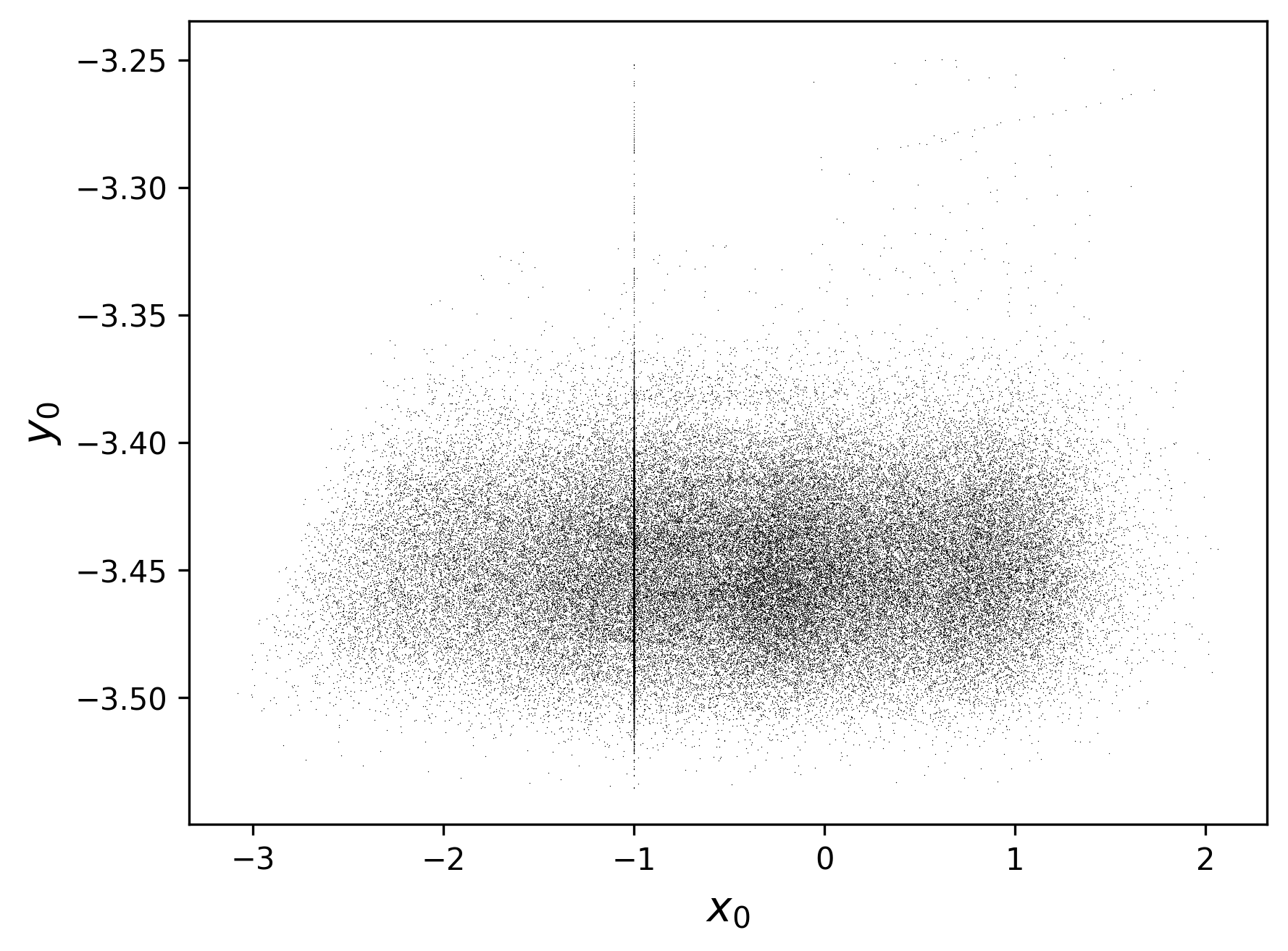} \label{fig:ring-slice-g1}}
    \end{adjustwidth}
    \caption{Projections of attractors of the 60-dimensional ring lattice system onto the $(x_0,y_0)$ plane for the homogeneous case, with~$x_{i,0}\in (-1,1)$, $y_{i,0}=-3.25$, $\sigma_i = -0.5$, and~$\alpha_i = 4.5$. The~attractors are plotted using orbits of length 100,000, and~we show attractors with the same four coupling strength values as in Figure~\ref{fig:random_x_graphs}: $g=0$ (uncoupled nonchaotic spiking); (\textbf{b}) $g=0.05$ (unsynchronized chaotic spiking); (\textbf{c}) $g=0.25$ (synchronized chaotic bursting); and (\textbf{d}) $g=1$ (synchronized hyperchaos).}
    \label{fig:ring-slices}
\end{figure}

In Section~\ref{sec:dynamics}, it was found that the three regimes of the ring lattice system of nonchaotic Rulkov neurons nearly always exhibit chaotic dynamics with positive maximal Lyapunov exponents. Therefore, it can be concluded that this system usually evolves towards some chaotic attractor in 60-dimensional state space. In~Figure~\ref{fig:ring-slices}, we plot projections of four attractors representative of the four dynamical regimes of the homogeneous case onto the $(x_0,y_0)$ plane. In~the uncoupled regime (Figure~\ref{fig:ring-slices}a), the~orbit is nonchaotic and periodic, so the attractor is composed of a finite number of isolated points. However, in~the coupled regimes (Figure \ref{fig:ring-slices}b--d), the~chaotic orbits produce much more complex attractors that appear fractal and strange. Although~these projections provide a simplified picture of the geometry of the attractors, they do not capture the full geometry of these objects embedded in 60-dimensional space. Thus, in~this section, we will focus on analyzing the geometry of these strange attractors by approximating their fractal~dimensions.

The fractal dimension serves as a critical tool for quantifying the geometric complexity of chaotic attractors. While Lyapunov exponents measure how sensitive the system is to initial conditions, the~fractal dimension characterizes how much of the state space is effectively explored by the system over time. In~particular, a~higher fractal dimension implies that the dynamics occupy a larger portion of the state space, potentially corresponding to a greater number of active degrees of freedom. This is especially important in high-dimensional coupled systems like the one studied in this paper, where complex collective behavior can arise from interactions among individually simple units. Moreover, comparing fractal dimensions across different coupling strengths and heterogeneity regimes allows us to investigate how different dynamical effects, such as synchronization, influence the \mbox{structure of the~attractor.}

To compute the basic box-counting dimension $d$ of a geometrical object, $n$-dimensional state space is covered with $n$-dimensional boxes of side length $\epsilon$. Then, the~number of boxes that the object touches, denoted as $N(\epsilon)$, is counted. Given this, the~relation
\begin{equation}
    N(\epsilon) \sim \epsilon^{-d}
    \label{eq:nep}
\end{equation}
is expected to hold~\cite{theiler}. However, an~issue immediately arises in the numerical computation of the fractal dimensions of attractors embedded in high-dimensional space. To~illustrate this problem, consider a 60-dimensional cube (which has fractal dimension $d=60$) filling some region of 60-dimensional space. If~we are to consider boxes with side length $\epsilon = \ell, \ell/2, $ and $\ell/4$, Equation~\eqref{eq:nep} indicates that $N(\ell/4)/N(\ell) = 4^{60}\approx 1.3\times 10^{36}$. Therefore, in the case of an attractor, it is necessary to sample at least in the order of $10^{36}$ points to obtain an accurate result for the fractal dimension of the attractor in this simplified case. This is clearly not feasible, so we turn to the Kaplan--Yorke conjecture to provide a computationally efficient approximation for the fractal dimensions of the ring system's~attractors.

The Kaplan--Yorke conjecture asserts that the Lyapunov spectrum of the orbit on an attractor is directly related to the attractor's dimension~\cite{kaplan-yorke}. Assuming that the Lyapunov spectrum is ordered from greatest to least, let $\kappa$ be the largest index such that
\begin{equation}
    \sum_{i=1}^{\kappa}\lambda_i\geq 0.
    \label{eq:kappa}
\end{equation}

Then, the~Lyapunov dimension $d_l$ is defined as
\begin{equation}
    d_l = \kappa + \frac{1}{|\lambda_{\kappa+1}|}\sum_{i=1}^{\kappa}\lambda_i.
    \label{eq:dl}
\end{equation}

The Kaplan--Yorke conjecture states that the Lyapunov dimension of an attractor is equal to its true fractal dimension $d$ \cite{nichols}.

\begin{figure}[H]
    \begin{adjustwidth}{-\extralength}{0cm}
    \centering
    \subfloat[\centering]{\includegraphics[scale=0.55]{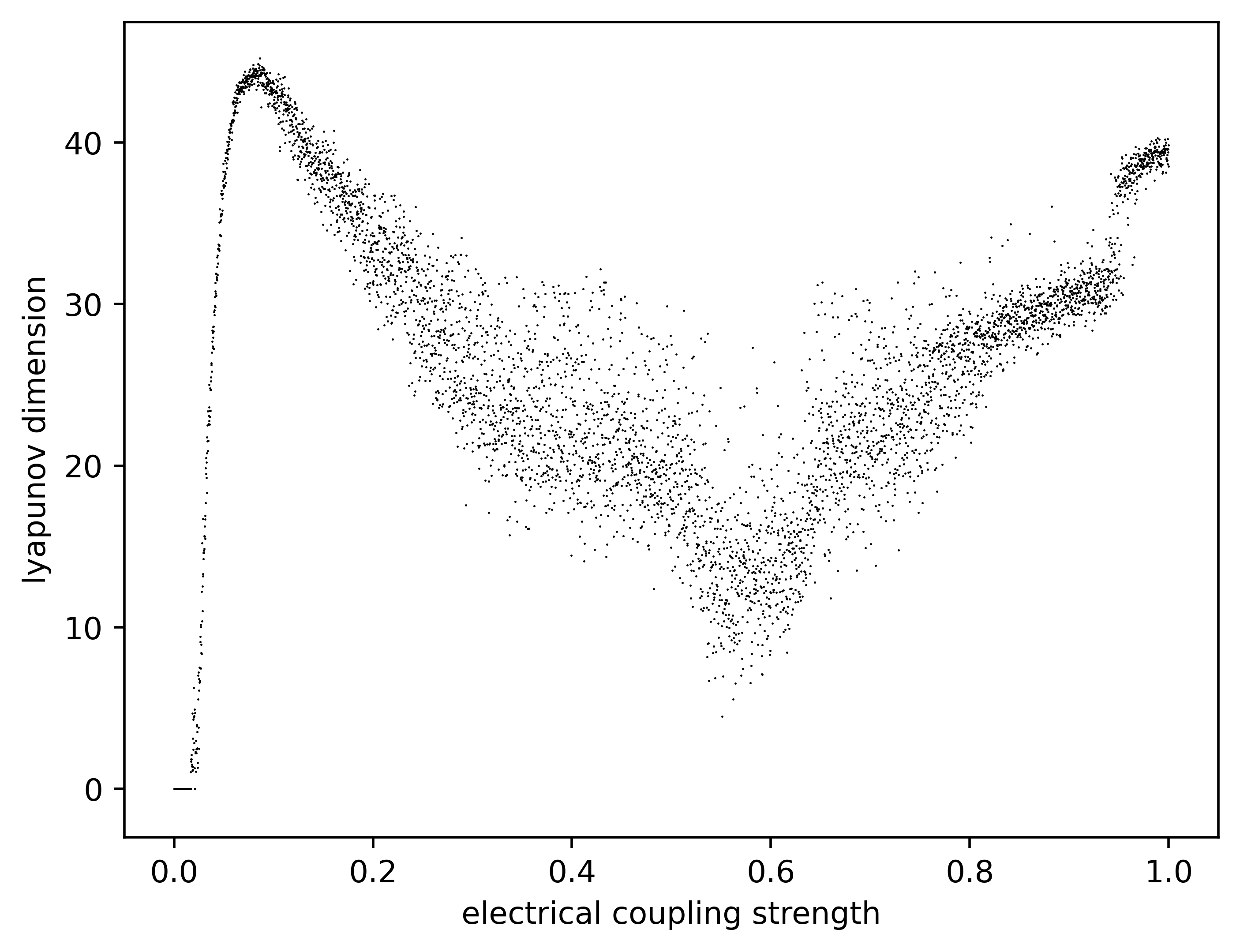} \label{fig:dl-random-x}} \\
    \subfloat[\centering]{\includegraphics[scale=0.55]{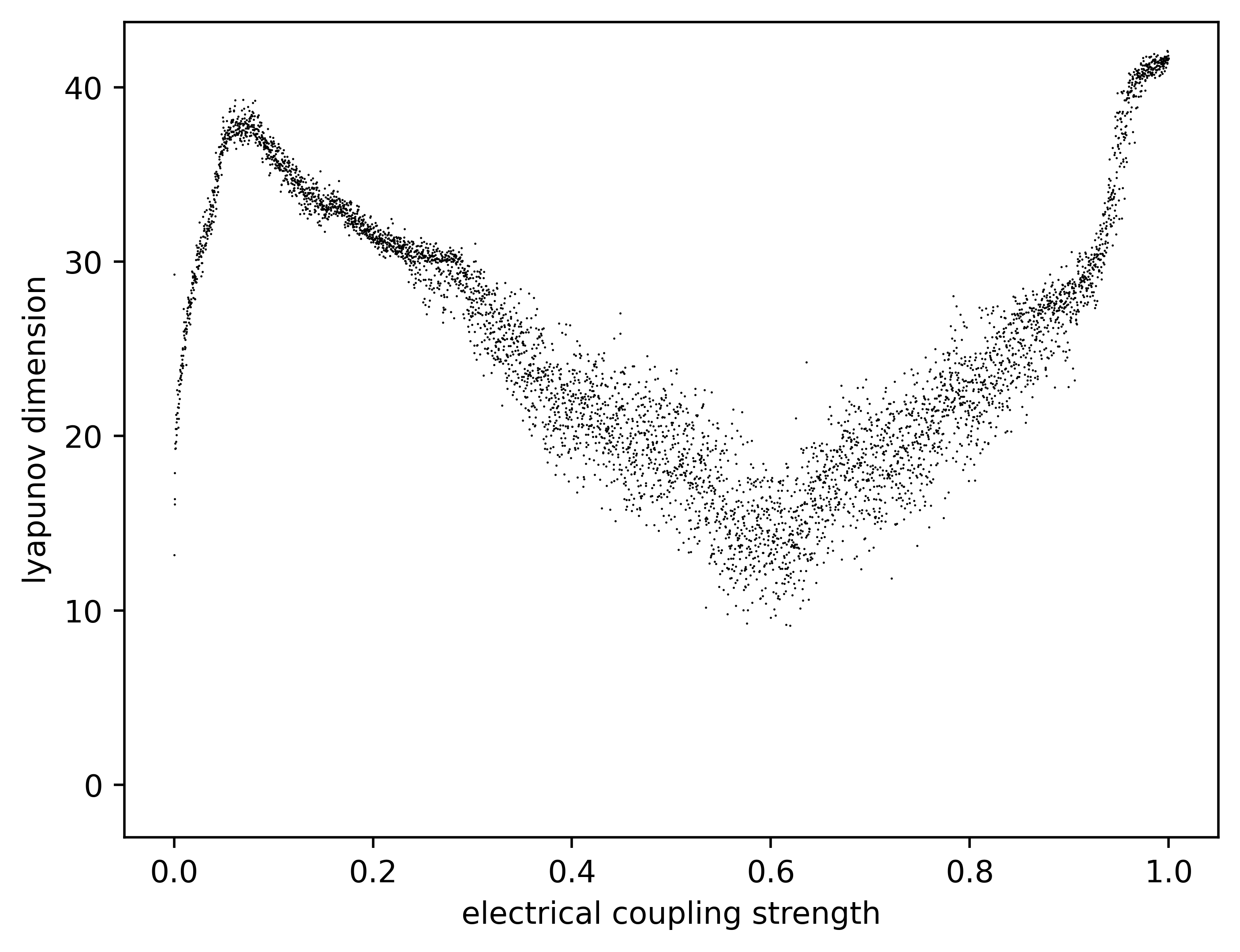} \label{fig:dl-random-sigma}}
    \subfloat[\centering]{\includegraphics[scale=0.55]{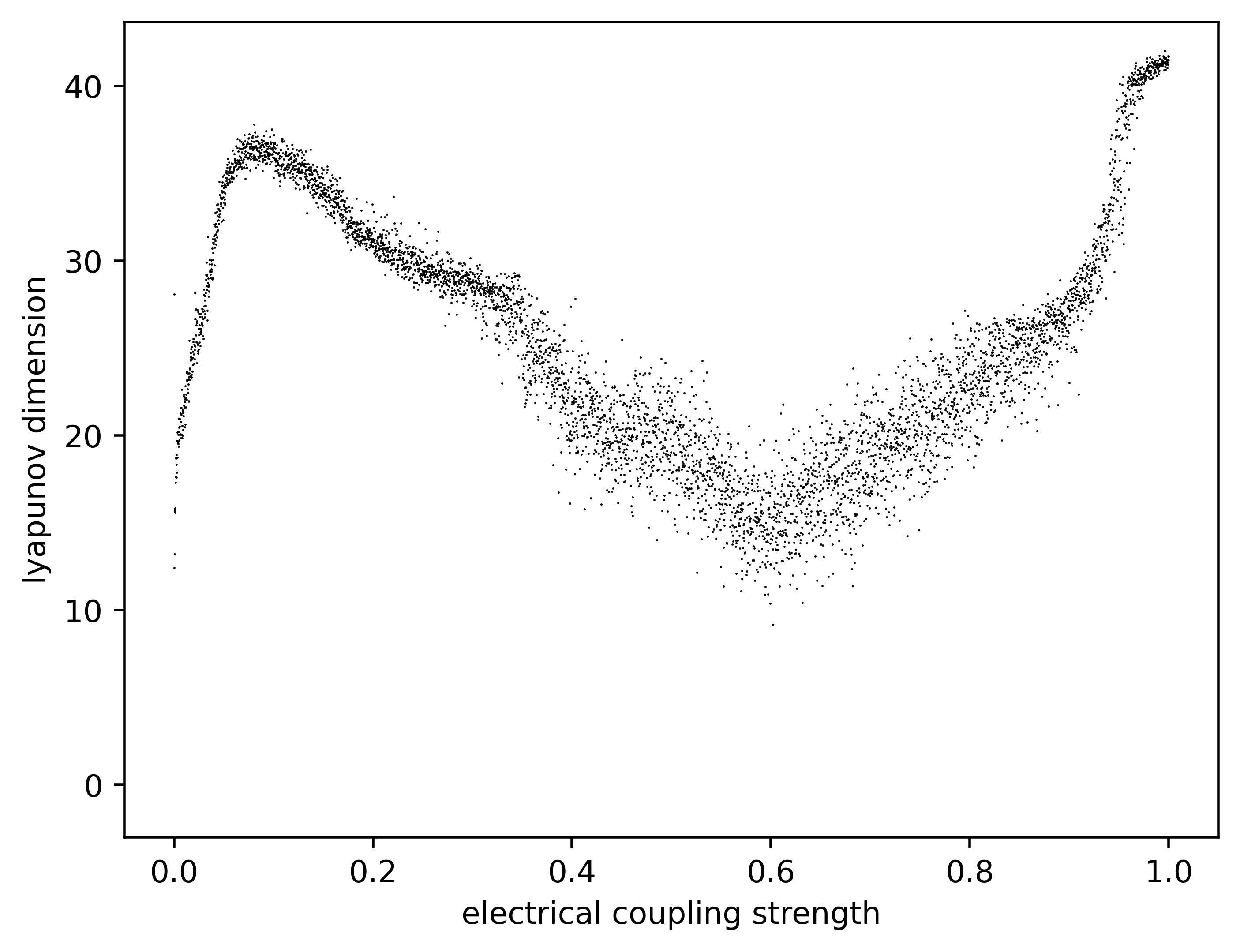} \label{fig:dl-random-alpha}}
    \end{adjustwidth}
    \caption{Graphs of the Lyapunov dimension $d_l$ against the electrical coupling strength $g$ for the (\textbf{a}) homogeneous case, (\textbf{b}) partially heterogeneous case, and~(\textbf{c}) fully heterogeneous case of the ring lattice system of $\zeta=30$ electrically coupled Rulkov neurons. The~Lyapunov dimension graphs show the same distinct regimes of behavior as the maximal Lyapunov exponent graphs in\linebreak   Figures~\ref{fig:max_lyap_exp_graph_random_x} and \ref{fig:ring-max-lyap-graphs-rand-sigandalph}, but~the Lyapunov dimension and maximal Lyapunov exponent graphs show different trends, as~described and explained in the text. The~Lyapunov spectra are calculated using orbits of length 1000, which is sufficient for convergence, and~the Lyapunov dimensions are computed using Equation \eqref{eq:dl}, which provides an accurate approximation for the true fractal dimensions of the~attractors.}
    \label{fig:dl-graphs}
\end{figure}

Although the Kaplan--Yorke conjecture remains unproven, it is well established that it holds in almost all cases~\cite{farmer}. However, we would still like to check for its validity in this system. Using the full Lyapunov spectra we computed in Section~\ref{sec:dynamics}, the~Lyapunov dimensions of the system can be calculated using Equations~\eqref{eq:kappa} and \eqref{eq:dl}. Then, graphs similar to the ones in Figures~\ref{fig:max_lyap_exp_graph_random_x} and \ref{fig:ring-max-lyap-graphs-rand-sigandalph} can be made by plotting the values of $d_l$ for many different values of $g$, which is displayed in Figure~\ref{fig:dl-graphs}. For~select values of $g$ in the homogeneous regime of the system, we also estimate the true fractal dimensions $d$ of the attractors with significant computation and careful application of Equation~\eqref{eq:nep}. Specifically, points are sampled on the attractors by generating many orbits of length $10^7$ for a given value of $g$, and close values of $\epsilon$ are chosen, where the sampled points scale according to their attractor. Then, a~linear regression is performed on $\ln N(\epsilon)$ vs. $\ln(1/\epsilon)$ and the slope is taken to be an approximation for $d$. The~results of this analysis are displayed in Table~\ref{tab:dim-compar}, where it is clear that the Lyapunov dimension $d_l$ falls well within a 5\% error of the estimated box-counting dimension $d$. Within~the margin of error in computing the Lyapunov spectrum of orbits on the system's attractors and box counting on the attractors, this indicates that the Kaplan--Yorke conjecture does hold for this system, so the Lyapunov dimensions will be used as an accurate approximation of the true fractal dimensions of the attractors. This enables us to investigate how the attractor dimensionality evolves with coupling strength across all three regimes of the ring lattice system without exponentially infeasible box counting in high-dimensional space, revealing deep connections between synchronization, chaos, and~the underlying geometric complexity of the system’s~dynamics.

\begin{table}[H]
    
    \caption{Comparisons between the Lyapunov dimension $(d_l)$ and estimated box-counting dimension $(d)$ of the chaotic attractors for select values of $g$ in the homogeneous regime of the ring lattice system. The~similarity between $d_l$ and $d$ suggests that the Kaplan--Yorke conjecture holds for this ring lattice system, so the Lyapunov dimension can be used as an accurate approximation for the true fractal~dimension. \label{tab:dim-compar}}
    \begin{tabularx}{\textwidth}{CCCC}
        \toprule
       \boldmath{ $g$} & \boldmath{$d_l$} & \boldmath{$d$} \textbf{(Estimated)} & \textbf{\% Error }\\
        \midrule
        0.1 & 43.27 & 42.86 & 0.96\% \\
        0.3 & 23.24 & 23.23 & 0.04\% \\
        0.6 & 15.80 & 16.23 & 2.65\% \\
        0.9 & 30.53 & 30.08 & 1.50\% \\
        \bottomrule
    \end{tabularx}
\end{table}
\unskip

In Figure~\ref{fig:dl-graphs}, it is immediately clear that all the chaotic attractors of the three regimes of the ring system are fractal, since their dimensions are spread out among different real values, not sticking to any defined integers. The~only true integer dimensions in these graphs are on the very left of Figure~\ref{fig:dl-graphs}a, where there are some attractors that have dimension 0. These are associated with the nonchaotic periodic orbit attractors on the left of Figure~\ref{fig:max_lyap_exp_graph_random_x}, which consist of a finite number of zero-dimensional points. One example of these orbits is displayed in the regular spiking of Figure~\ref{fig:random_x_graphs}a. Another notable observation is that these attractors take up a large number of dimensions of state space. Because~the state space of this system is so large, we might expect the attractors to take up only a small number of its dimensions, but~instead, the~strange attractors take up a substantial number of them for many values of $g$, with~some of the largest of these attractors taking up close to 45 of the 60 total~dimensions. 

Comparing Figure~\ref{fig:dl-graphs} to the graphs of $\lambda_1$ vs. $g$ in Figures~\ref{fig:max_lyap_exp_graph_random_x} and \ref{fig:ring-max-lyap-graphs-rand-sigandalph}, it can be seen that the Lyapunov dimension $d_l$ follows a similar pattern of increasing through the chaotic spiking domain, decreasing as the neurons start to burst in sync with each other, and then increasing again as synchronized hyperchaos is reached. This is to be expected, because the Lyapunov dimension is calculated directly from the set of Lyapunov exponents. There is also a similarity in how the $d_l$ and $\lambda_1$ values are distributed across the different regimes. Specifically, the~$\lambda_1$ values are more erratic and spread out in the homogeneous case than they are in the partially and fully heterogeneous cases, which is also reflected in the $d_l$ values to some degree. Namely, the~values of $d_l$ in Figure~\ref{fig:dl-graphs}a are more vertically spread out in the synchronized bursting domain. There are two main reasons for this difference in variability. The~first has to do with the discrete-time nature of the Rulkov model and is similar to the reason for the existence of complex multistability in the homogeneous synchronized bursting regime~\cite{rulkov}. Specifically, because~the system is governed by a discrete-time map, a~small variation in $g$ can cause the individually nonchaotic spikes of the homogeneous neurons to lock onto each other with different frequency ratios, leading to more variability in the dynamics of the homogeneous regime and the geometry of its attractors. The~second reason for this difference in variability has to do with the fact that the heterogeneous cases of the system contain individually silent and low-frequency bursting neurons (e.g., $\mb{x}_5$ and $\mb{x}_6$ in Figures~\ref{fig:random_sigma_graphs}a and \ref{fig:random_alpha_graphs}a), which lead to more nonchaotic behavior in the synchronized bursting regime (see $x_{5,k}$ and $x_{6,k}$ in Figures~\ref{fig:random_sigma_graphs}c and \ref{fig:random_alpha_graphs}c). This nonchaotic behavior contributes to the smaller upper bounds of the fractal dimensions in the heterogeneous regime compared to the homogeneous regime and,~hence, to lower~variability.

In addition to this, there are some very clear differences between the trends of the maximal Lyapunov exponent $\lambda_1$ and the Lyapunov dimension $d_l$. The~most apparent difference is in the peaks of the $\lambda_1$ vs. $g$ graphs and the $d_l$ vs. $g$ graphs, with~both peaks in both graphs being associated with chaotic spiking around $g=0.1$ and synchronized hyperchaos around $g=1$. In~the $\lambda_1$ vs. $g$ graphs, the~peak in the region of synchronized hyperchaos is always higher than the peak in the region of chaotic spiking, a~fact that is extremely apparent in Figure~\ref{fig:ring-max-lyap-graphs-rand-sigandalph} (the partially and fully heterogeneous cases), where the peaks on the right dwarf the peaks on the left. However, in~the graphs of $d_l$ vs. $g$, the~peaks are similar in height, and~in Figure~\ref{fig:dl-graphs}a (the homogeneous case), the~left peak is actually higher than the right peak. This means that, for~this system, the~chaotic spiking attractor that appears when the electrical coupling strength is relatively small has a higher fractal dimension than the attractor that appears when the electrical coupling strength is very large, which is a somewhat surprising~result. 

To explain these interesting phenomena, we will draw on the connection between dynamics and geometry posited by the Kaplan--Yorke conjecture. First, we address the dramatic difference in the heights of the left and right peaks when comparing the graphs of the maximal Lyapunov exponent and the fractal dimension (Figures~\ref{fig:max_lyap_exp_graph_random_x}, \ref{fig:ring-max-lyap-graphs-rand-sigandalph}, and~\ref{fig:dl-graphs}). In~the region of the right peak, the~neurons are exhibiting synchronized, strong chaos, whereas in the region of the left peak, the~neurons are exhibiting unsynchronized, weaker chaos. The~strength of the chaotic dynamics as a whole is reflected in the maximal Lyapunov exponent, demonstrated in Figure~\ref{fig:max_lyap_exp_graph_random_x} with a higher right peak and Figure~\ref{fig:ring-max-lyap-graphs-rand-sigandalph} with significantly higher right peaks. However, when considering the attractor dimensions, the~Kaplan--Yorke conjecture indicates that the entire Lyapunov spectrum must be considered. In~the region of the right peak, the~synchronized chaos is indicative of the strong chaotic dynamics being ``connected,'' or in the language of Lyapunov exponents, only a few eigenvectors of the Jacobian having positive eigenvalues. In~other words, perturbing the system along one of these chaotic directions, indicative of perturbing all of the neurons in the same way, will result in this perturbation growing, but~perturbing the system along any of the other directions will result in the perturbation shrinking due to the system falling back into synchronization. However, in~the region of the left peak, the~unsynchronized chaos is indicative of each neuron having its own ``direction'' of chaos, which results in many more positive Lyapunov exponents and, according to~the Kaplan--Yorke conjecture, a~higher fractal dimension. For~example, in~the homogeneous case, the~system has 9 positive Lyapunov exponents for $g=0.95$ and 18 positive Lyapunov exponents for $g=0.1$. This explains the dramatic increase in the left peaks compared to the right peaks when comparing Figure~\ref{fig:dl-graphs} to Figures~\ref{fig:max_lyap_exp_graph_random_x} and \ref{fig:ring-max-lyap-graphs-rand-sigandalph}. The~differing number of positive Lyapunov exponents also provides a mathematical justification for referring to the various dynamical regimes as synchronized (low number) or unsynchronized (high number). This connection between the number of positive Lyapunov exponents and synchrony is well-established in the literature~\cite{pecora, lai}.

The reversed peak height difference in the homogeneous case's fractal dimension graph (Figure~\ref{fig:dl-graphs}a) compared to the heterogeneous cases (Figure~\ref{fig:dl-graphs}b,c) can be explained in a similar manner to the aforementioned variability discrepancy between the homogeneous and heterogeneous cases. Specifically, the~left peaks of the heterogeneous cases' fractal dimension graphs are lowered due to the contribution from individual silent and low-frequency bursting neurons, resulting in nonchaotic behavior. However, in~the region of the right peak, the~effect of heterogeneity is reversed, namely, the~right peak is raised because the variations in the individual neurons' parameters results in greater sensitivity to perturbations and stronger chaotic dynamics in the synchronized hyperchaotic domain (see the right peaks of Figure~\ref{fig:ring-max-lyap-graphs-rand-sigandalph} vs. Figure~\ref{fig:max_lyap_exp_graph_random_x}), increasing the magnitude of the positive Lyapunov exponents and the fractal dimension. This analysis makes it clear that although the maximal Lyapunov exponent quantifies how chaotic the dynamics on the ring lattice attractors are as a whole, it does not directly correlate to the attractors' dimensionality or strangeness. To achieve this, as~the Kaplan--Yorke conjecture indicates, we need the entire Lyapunov spectrum, which captures more information about the collective dynamics and individual behavior of the coupled~neurons.

\section{Conclusions}
\label{sec:conclusions}

In this paper, the~dynamics and geometry that emerge from a model consisting of a ring of electrically coupled nonchaotic Rulkov neurons were investigated. Extensive numerical simulations were performed to analyze the dynamics of homogeneous, partially heterogeneous, and~fully heterogeneous regimes of a ring lattice system of $\zeta=30$ neurons. It was found that a variety of chaotic behaviors emerged from individually nonchaotic neurons, including chaotic spiking, synchronized chaotic bursts, and~synchronized hyperchaos. To~quantify the chaos of the ring system, its $2\zeta \times 2\zeta$ Jacobian matrix was calculated, and~its maximal Lyapunov exponents were computed for a range of electrical coupling strengths. Using the QR factorization method for computing Lyapunov spectra, the~fractal dimensions of the attractors in 60-dimensional state space were approximated via the Kaplan--Yorke conjecture. It was found that all chaotic attractors across the three regimes were fractal, and~that for certain coupling strengths, the~attractors occupied significant portions of the 60-dimensional state space. When comparing the Lyapunov dimensions of the ring lattice system to its maximal Lyapunov exponents, it was observed that although both quantities followed a similar pattern of increasing and decreasing with varying coupling strength, they were not directly correlated, reflecting more subtle emergent behaviors due to electrical coupling and the complex relationship between dynamics and geometry posited by the Kaplan--Yorke conjecture. These findings provide a deeper understanding of how complexity can emerge in coupled networks and may inform future studies of synchronization, information flow, and~pattern formation in biological and artificial~systems.

Looking ahead, the~calculation of the complex Jacobian matrix of the ring model can be naturally extended to more complex lattices of neurons, such as a mesh, torus, or~sphere, as~well as an all-to-all coupled system. Although~these have been studied in the context of a mean field of chaotic Rulkov neurons~\cite{rulkov2}, such studies have never been conducted with the more experimentally applicable electrical coupling of Rulkov neurons, to the best of our knowledge. With~more current connections, more interesting hyperchaotic dynamics are likely to appear. In~future works, we will investigate the dynamical and geometrical properties of $N$-dimensional lattices of electrically coupled nonchaotic Rulkov neurons with periodic boundary conditions~\cite{n-dim}. Beyond~numerical simulation, a~rigorous, theoretical analysis of this system should be conducted in future work. Additionally, future research aimed at validating the model's complex dynamics through physical systems is also encouraged. One promising direction involves digital hardware validation, such as implementing the ring lattice model on FPGA (Field-Programmable Gate Array) platforms. As~demonstrated in recent studies~\cite{fpga-1, fpga-2, fpga-3}, FPGA-based realizations of neuron models can effectively capture nonlinear dynamics while offering practical advantages in terms of speed and reconfigurability. Such an approach would not only demonstrate the physical realizability of our model, but also open pathways toward engineering applications, such as neuromorphic computing or real-time signal generation using high-dimensional chaos. At~the same time, we suggest further investigation into the biological relevance of these findings by attempting to observe similar collective dynamics in real neuronal systems. This work builds on existing experimental studies of coupled biological neurons~\cite{elson, abarbanel, varona, okun, pendeliuk} and could help determine whether similar complex dynamical regimes and geometrical structures can be identified in real neural tissue or~cultures.






\funding{This research received no external~funding.}

\dataavailability{The original data presented in the study are openly available in a
GitHub repository at \url{https://github.com/brandon-bd-le/RulkovRing2025} (accessed on 2 September 2025).}

\acknowledgments{The author thanks Nivika A. Gandhi and Mark S. Hannum for~engaging in discussions.}

\conflictsofinterest{The author declares no conflicts of~interest.} 



\appendixtitles{yes} 
\appendixstart
\appendix

\section{Jacobian~Matrix}
\label{appen:jacobian}

In this appendix, we outline a sketch of the derivation of the $2\zeta\times 2\zeta$ Jacobian matrix of a Rulkov ring lattice system governed by the iteration function in Equation~\eqref{eq:ring-iteration-func}. For~a full detailed derivation of the Jacobian, see Section~7.2 of Ref.~\cite{bn}. Here, we derive the $mp$th entry of $J(\mathbf{X})$:
\begin{equation}
    J_{mp}(\mathbf{X}) = \frac{\partial F\e{m}}{\partial X\e{p}}.
    \label{eq:compactified-ring-jacobian}
\end{equation}

From Equation~\eqref{eq:ring-system-state-vector}, it is clear that when $p$ is odd, we are differentiating with respect to the fast variable of the neuron with index $i=(p-1)/2$, and~when $p$ is even, we are differentiating with respect the slow variable of the neuron with index $i=p/2-1$. Similarly, from~Equation~\eqref{eq:ring-iteration-func}, when $m$ is odd, we are differentiating the piecewise fast-variable function $f$ of the neuron with index $i=(m-1)/2$, and~when $m$ is even, we are differentiating the slow-variable function of the neuron with index $i=m/2-1$. 
\pagebreak

Let us first consider even $m$, or~$m\bmod 2 = 0$. According to~Equations~\eqref{eq:rulkov_coupled_mapping} and \eqref{eq:ring-coup-params}, the~slow-variable iteration function for neuron $i=m/2-1$ is
\begingroup\makeatletter\def\f@size{9}\check@mathfonts
\def\maketag@@@#1{\hbox{\m@th\normalsize \normalfont#1}}%
\begin{equation}
    F\e{m} = y_{m/2-1}-\mu x_{m/2-1} + \mu\Bigg[\sigma_{m/2-1} + \frac{g}{2}(x_{(m/2-2)\bmod\zeta} + x_{(m/2)\bmod\zeta} - 2x_{m/2-1})\Bigg].
    \label{eq:even-m-ring-function}
\end{equation}
\endgroup

This function only depends on $y_{m/2-1}$, $x_{m/2-1}$, $x_{(m/2-2)\bmod\zeta}$, and~$x_{(m/2)\bmod\zeta}$, so the derivative with respect to any other variable will vanish. Therefore, we need only determine the values of $p$ that will make $X\e{p}$ equal one of these variables that yields a non-vanishing derivative, where careful attention must be paid to the values of $m$ that are near the loop-around point of the~ring. 

For odd $m$ ($m\bmod 2 = 1$), we are differentiating the fast-variable iteration function of neuron $i = (m-1)/2$. Therefore, according to~Equations~\eqref{eq:rulkov_1_fast_equation}, \eqref{eq:rulkov_coupled_mapping}, and~\eqref{eq:ring-coup-params}, 
\vspace{-8pt}
\begingroup\makeatletter\def\f@size{9}\check@mathfonts
\def\maketag@@@#1{\hbox{\m@th\normalsize \normalfont#1}}%
\begin{adjustwidth}{-\extralength}{0cm}
\begin{equation}
    \begin{split}
        F\e{m} &= f(x_{(m-1)/2},y_{(m-1)/2} + \mathfrak{C}_{(m-1)/2};\alpha_{(m-1)/2}) \\[0.25cm]
        &= \begin{cases}
            \displaystyle \frac{\alpha_{(m-1)/2}}{1-x_{(m-1)/2}} + y_{(m-1)/2} + \frac{g}{2}(x_{[(m-3)/2] \bmod\zeta} + x_{[(m+1)/2] \bmod\zeta} - 2x_{(m-1)/2}), & x_{(m-1)/2} \leq 0 \vspace{0.5cm} \\
            \displaystyle \alpha_{(m-1)/2} + y_{(m-1)/2} + \frac{g}{2}(x_{[(m-3)/2] \bmod\zeta} + x_{[(m+1)/2] \bmod\zeta} - 2x_{(m-1)/2}), & 0 < x_{(m-1)/2} < \alpha_{(m-1)/2} \\
             & + y_{(m-1)/2} + \mathfrak{C}_{(m-1)/2} \vspace{0.5cm} \\
            -1, & x_{(m-1)/2} \geq \alpha_{(m-1)/2} \\
            & + y_{(m-1)/2} + \mathfrak{C}_{(m-1)/2} \\
        \end{cases}.
    \end{split}
    \label{eq:piecewise-ring-function}
\end{equation}
\end{adjustwidth}
\endgroup
In the case where $x_{(m-1)/2} \leq 0$, the~only variables present are $y_{(m-1)/2}$, $x_{(m-1)/2}$, $x_{[(m-3)/2]\bmod\zeta}$, and~$x_{[(m+1)/2]\bmod\zeta}$, so we can systematically determine the values of $p$ that yield non-zero derivatives in a similar fashion to the odd $m$ function. In~the case where $0 < x_{(m-1)/2} < \alpha_{(m-1)/2} + y_{(m-1)/2} + \mathfrak{C}_{(m-1)/2}$, we have different non-zero derivatives, since the function piece is different, but~this piece depends on the same variables as the first piece, so the same relevant $p$ values apply. In~the case where $x_{(m-1)/2} \geq \alpha_{(m-1)/2} + y_{(m-1)/2} + \mathfrak{C}_{(m-1)/2}$, the~derivative with respect to any variable is trivial. Putting all of this together yields the Jacobian entry $J_{mp}(\mathbf{X})$ central to the Lyapunov spectrum calculation for a Rulkov ring lattice system:
\begingroup\makeatletter\def\f@size{9}\check@mathfonts
\def\maketag@@@#1{\hbox{\m@th\normalsize \normalfont#1}}%
\begin{adjustwidth}{-\extralength}{0cm}
\begin{equation}
    J_{mp}(\mathbf{X}) = 
    \begin{cases}
        \begin{cases}
        
            \begin{cases}
                1, & \hspace{0.325cm}\text{if }p=m+1, \\
                \displaystyle \frac{\alpha_{(m-1)/2}}{(1-x_{(m-1)/2})^2} - g, & \hspace{0.325cm}\text{if }p=m,\\
                 g/2, & \begin{cases}
                    \text{if }p=m-2,\\
                    \phantom{if }\text{ and }m\neq 1, \\
                    \text{or }p=2\zeta-1,\\
                    \phantom{if }\text{ and }m=1, \\
                    \text{or }p=m+2,\\
                    \phantom{if }\text{ and }m\neq 2\zeta - 1, \\
                    \text{or }p=1,\\
                    \phantom{if }\text{ and }m=2\zeta - 1,
                \end{cases} \\
                0, & \hspace{0.325cm}\text{otherwise,}
            \end{cases} \text{for } x_{(m-1)/2} \leq 0, \\
            \\
            \\
            \begin{cases}
                1, & \hspace{2.225cm}\text{if }p=m+1, \\
                - g, & \hspace{2.225cm}\text{if }p=m,\\
                g/2, & \hspace{1.9cm}\begin{cases}
                    \text{if }p=m-2,\\
                    \phantom{if }\text{ and }m\neq 1, \\
                    \text{or }p=2\zeta-1,\\
                    \phantom{if }\text{ and }m=1, \\
                    \text{or }p=m+2,\\
                    \phantom{if }\text{ and }m\neq 2\zeta - 1, \\
                    \text{or }p=1,\\
                    \phantom{if }\text{ and }m=2\zeta - 1,
                \end{cases} \stackunder{\text{for $0 < x_{(m-1)/2} < \alpha_{(m-1)/2}$}}{\text{$+ y_{(m-1)/2} + \mathfrak{C}_{(m-1)/2},$}} \\
                0, & \hspace{2.225cm}\text{otherwise,}
            \end{cases}\\
            \\
            \\
            0, \hspace{6.685cm} \text{for} \stackunder{x_{(m-1)/2} \geq \alpha_{(m-1)/2}}{+ y_{(m-1)/2} + \mathfrak{C}_{(m-1)/2},}\\
            
        \end{cases} & \text{when } m\bmod 2=1 \\
        \\
        \\
        \begin{cases}
            1, & \hspace{0.67cm}\text{if }p=m \\
            -\mu(1+g), & \hspace{0.67cm}\text{if }p=m-1 \\
            \mu g/2, & \hspace{0.345cm}
            \begin{cases}
                \text{if }p=m-3\\
                    \phantom{if }\text{ and }m\neq 2 \\
                \text{or }p=2\zeta-1\\
                    \phantom{if }\text{ and }m=2 \\
                \text{or }p=m+1\\
                    \phantom{if }\text{ and }m\neq 2\zeta \\
                \text{or }p=1\\
                    \phantom{if }\text{ and }m=2\zeta
            \end{cases} \\
        0, & \hspace{0.67cm}\text{otherwise,}
        \end{cases} & \text{when } m\bmod 2=0
    \end{cases}.
    \label{eq:THE-jacobian-entry}
\end{equation}
\end{adjustwidth}
\endgroup

\section{QR Factorization Method of Lyapunov Spectrum~Calculation}
\label{appen:qr}

In this appendix, we detail how the Lyapunov spectrum $\{\lambda_1,\lambda_2,\hdots,\lambda_n\}$ is computed using the QR factorization method from Ref.~\cite{eckmann}, which utilizes the Jacobian matrix $J(\mb{X})$ we derived in Appendix \ref{appen:jacobian}. Here, we follow the derivation of this algorithm in Ref.~\cite{brandon}.

To begin, we perform a QR decomposition on $J(\mb{X}_0)$, denoting
\begin{equation}
    J(\mb{X}_0) = Q^{(1)}R^{(1)}.
    \label{eq:decomp-1}
\end{equation}

Then, for~$k=2,3,\hdots,t$, we recursively define
\begin{equation}
    J(\mb{X}_{k-1})Q^{(k-1)} = J_k^*
\end{equation}
and decompose the matrix $J_k^*$ into
\begin{equation}
    J_k^* = Q^{(k)}R^{(k)}.
    \label{eq:decomp-2}
\end{equation}

It follows from this that any $J(\mb{X}_{k-1})$ can be written as $J(\mb{X}_{k-1}) = Q^{(k)}R^{(k)}(Q^{(k-1)})^\top$, so we can write
\begin{equation}
    \begin{split}
        J^t &= J(\mb{X}_{t-1})J(\mb{X}_{t-2})\cdots J(\mb{X}_0) = Q^{(t)}R^{(t)}R^{(t-1)}\cdots R^{(1)} \\
        &= Q^{(t)}\Upsilon^{(t)},
    \end{split}
    \label{eq:jt}
\end{equation}
where $\Upsilon^{(t)} = R^{(t)}R^{(t-1)}\cdots R^{(1)}$ is an upper triangular~matrix.

For some initial state $\mb{X}_0$ and small perturbation in the direction of the unit vector $\mb{U}_0$, the~associated Lyapunov exponent is~\cite{sandri}
\begin{equation}
    \lambda = \lim_{t\to\infty}\frac{1}{t}\ln|J^t\mb{U}_0|.
\end{equation}

Substituting in Equation~\eqref{eq:jt}, taking $\mb{U}_0$ to be a normalized eigenvector of $\Upsilon^{(t)}$, and~using the orthogonality of $Q^{(t)}$ yields
\begin{equation}
    \lambda_i = \lim_{t\to\infty}\frac{1}{t}\ln|Q^{(t)}\upsilon_{ii}^{(t)}\mb{U}_0| = \lim_{t\to\infty}\frac{1}{t}\ln|\upsilon_{ii}^{(t)}|,
    \label{eq:upsilon}
\end{equation}
where $\upsilon_{ii}^{(t)}$ is the $ii$th entry of $\Upsilon^{(t)}$, and we arrange the column vectors of the $Q$ matrices so that the diagonal entries of their associated $R$ matrices are ordered from greatest to least. According to~the definition of $\Upsilon^{(t)}$, we determine that $\upsilon_{ii}^{(t)} = r_{ii}^{(t)}r_{ii}^{(t-1)}\cdots r_{ii}^{(1)}$, where $r_{ii}^{(k)}$ is the $ii$th entry of $R^{(k)}$. Then, we can write Equation~\eqref{eq:upsilon} in the computationally efficient form
\begin{equation}
    \lambda_i = \lim_{t\to\infty}\frac{1}{t}\sum_{k=1}^t\ln|r_{ii}^{(k)}|.
    \label{eq:lyap-exps}
\end{equation}

Using this method, by~choosing a large value of $t$, we can compute the $2\zeta$ Lyapunov exponents of a Rulkov ring lattice system by using Equation~\eqref{eq:THE-jacobian-entry} to calculate the Jacobian matrices $J(\mb{X})$ for all $\mb{X}\in\{\mb{X}_0,\mb{X}_1,\hdots,\mb{X}_{t-1}\}$, performing the decompositions in\linebreak   Equations~\eqref{eq:decomp-1} and \eqref{eq:decomp-2}, and~plugging the diagonal entries of the resulting $R$ matrices into Equation~\eqref{eq:lyap-exps}.

Computing the full Lyapunov spectrum for a high-dimensional system involves a significant computational burden, primarily due to the need for repeated QR factorizations of the Jacobian matrix at each time step. For~a system of dimension $n$, QR decomposition requires $\mathcal{O}(n^3)$ operations per step, and~the total cost scales linearly with the number of steps in the simulation. In~our case, the~60-dimensional state space involves evolving and orthonormalizing 60 perturbation vectors throughout the simulation, resulting in substantial memory use and computational effort. This overhead is compounded by the fact that QR factorization is not trivially parallelizable, making it one of the dominant contributors to total runtime in high-dimensional Lyapunov~analysis.

\section{Pseudocode}
\label{appen:pseudocode}

\begin{algorithm}[H]
\caption{Single Rulkov~Neuron}
\begin{algorithmic}[1]
\Function{FastRulkovMap}{$x, y, \alpha$}
    \If{$x \le 0$}
        \State \Return $\alpha / (1 - x) + y$
    \ElsIf{$0 < x < \alpha + y$}
        \State \Return $\alpha + y$
    \Else
        \State \Return $-1$
    \EndIf
\EndFunction
\Function{RulkovMap}{$x, y, \sigma, \alpha, \mu$}
    \State $x^{+} \gets$ \Call{FastRulkovMap}{$x, y, \alpha$}
    \State $y^{+} \gets y - \mu \cdot (x - \sigma)$
    \State \Return $(x^{+}, y^{+})$
\EndFunction
\end{algorithmic}
\end{algorithm}
\unskip
\begin{algorithm}[H]
\caption{Coupled Rulkov Ring~Update}
\begin{algorithmic}[1]
\Function{RingCouplingX}{$X, i, g$}
    \State $left \gets (i-1) \bmod \zeta$
    \State $right \gets (i+1) \bmod \zeta$
    \State \Return $(g/2) \cdot (x_{left} + x_{right} - 2 \cdot x_i)$
\EndFunction
\Function{CoupledUpdate}{$X, \sigma, \alpha, \mu, g$}
    \For{$i = 0$ to $\zeta-1$}
        \State $C \gets$ \Call{RingCouplingX}{$X, i, g$}
        \State $(x_i, y_i) \gets X[i]$
        \State $x_i^{+} \gets$ \Call{FastRulkovMap}{$x_i, y_i + C, \alpha[i]$}
        \State $y_i^{+} \gets y_i - \mu \cdot x_i + \mu \cdot (\sigma[i] + C)$
        \State $X^{+}[i] \gets (x_i^{+}, y_i^{+})$
    \EndFor
    \State \Return $X^{+}$
\EndFunction
\end{algorithmic}
\end{algorithm}
\unskip
\begin{algorithm}[H]
\caption{Orbit~Generation}
\begin{algorithmic}[1]
\Function{GenerateRingOrbit}{$X_0, \sigma, \alpha, \mu, g, \zeta, T$}
    \State $Traj \gets [X_0]$
    \State $X \gets X_0$
    \For{$t = 1$ to $T$}
        \State $X \gets$ \Call{CoupledUpdate}{$X, \sigma, \alpha, \mu, g$}
        \State Append $X$ to $Traj$
    \EndFor
    \State \Return $Traj$
\EndFunction
\end{algorithmic}
\end{algorithm}

\begin{algorithm}[H]
\caption{Lyapunov Spectrum via QR~Factorization}
\begin{algorithmic}[1]
\Function{LyapSpectrumQR}{$Jlist, \zeta$}
    \State $Q \gets I_{2\zeta}$; $sums \gets 0$
    \For{$J$ in $Jlist$}
        \State $A \gets J \cdot Q$
        \State $(Q, R) \gets$ QR-factorization of $A$
        \For{$j = 0$ to $2\zeta-1$}
            \State $sums[j] \gets sums[j] + \log(|R_{jj}|)$
        \EndFor
    \EndFor
    \State $\lambda \gets sums / |Jlist|$
    \State \Return $\lambda$ sorted in descending order
\EndFunction
\end{algorithmic}
\end{algorithm}
\unskip
\begin{algorithm}[H]
\caption{Kaplan--Yorke~Dimension}
\begin{algorithmic}[1]
\Function{KYDimension}{$\lambda$}
    \State $S \gets 0$
    \For{$k = 1$ to $2\zeta$}
        \State $S \gets S + \lambda_k$
        \If{$S \le 0$}
            \State $\kappa \gets k - 1$
            \State $S \gets S - \lambda_k$
            \State \textbf{break}
        \EndIf
    \EndFor
    \State \Return $\kappa - S / \lambda_{\kappa+1}$
\EndFunction
\end{algorithmic}
\end{algorithm}

\section{Random Initial States and~Parameters}
\label{appen:supp}

In all three regimes of the ring lattice system we studied, we used random initial states and parameters. In~this appendix, we list these random values for the sake of reproducibility of results. We use the notations $\boldsymbol{\alpha} = (\alpha_1,\hdots,\alpha_{\zeta})$ and $\boldsymbol{\sigma} = (\sigma_1,\hdots,\sigma_{\zeta})$

In all three cases, we use the initial state
\vspace{-12pt}
\begin{adjustwidth}{-\extralength}{0cm}
\begin{equation}
    \begin{split}
        \mathbf{X}_0 &= (0.68921784, -3.25, -0.94561073, -3.25, -0.95674631, -3.25,  0.91870134, -3.25, \\
         &\mathrel{\phantom{=}} -0.32012381, -3.25, -0.23746836, -3.25, -0.43906743, -3.25, -0.48671017, -3.25, \\
         &\mathrel{\phantom{=}} -0.37578533, -3.25, -0.00613823, -3.25, 0.25990663, -3.25, -0.54103868, -3.25, \\
         &\mathrel{\phantom{=}} 0.12110471, -3.25,  0.71202085, -3.25, 0.689336, -3.25,   -0.03260047, -3.25, \\
         &\mathrel{\phantom{=}} -0.90907325, -3.25,  0.93270227, -3.25, 0.51953315, -3.25, -0.46783677, -3.25, \\
         &\mathrel{\phantom{=}} -0.96738424, -3.25, -0.50828432, -3.25, -0.60388469, -3.25, -0.56644705, -3.25, \\
         &\mathrel{\phantom{=}} -0.42772621, -3.25,  0.7716625, -3.25, -0.60336517, -3.25,  0.88158364, -3.25, \\
         &\mathrel{\phantom{=}} 0.0269842, -3.25,   0.42512831, -3.25),
    \end{split}
    \label{eq:big-initial-state}
\end{equation}
\end{adjustwidth}
with $x_{i,0}\in(-1,1)$. In~the partially and fully heterogeneous cases, we use the $\boldsymbol{\sigma}$ vector
\vspace{-12pt}
\begin{adjustwidth}{-\extralength}{0cm}
\begin{equation}
    \begin{split}
        \boldsymbol{\sigma} &= ( -0.63903048, -0.87244087, -1.16110093, -0.63908737, -0.73103576, -1.23516699, \\
        &\mathrel{\phantom{=}} -1.09564519, -0.57564289, -0.75055299, -1.01278976, -0.61265545, -0.75514189, \\
        &\mathrel{\phantom{=}} -0.89922568, -1.24012127, -0.87605023, -0.94846269, -0.78963971, -0.94874874, \\
        &\mathrel{\phantom{=}} -1.31858036, -1.34727902, -0.7076453, -1.10631486, -1.33635792, -1.48435264, \\
        &\mathrel{\phantom{=}} -0.76176103, -1.17618267, -1.10236959, -0.66159308, -1.27849639, -0.9145025 ),
    \end{split}
    \label{eq:big-sigma-vector}
\end{equation}
\end{adjustwidth}
with $\sigma_i\in (-1.5,-0.5)$. In~the fully heterogeneous case, we use the $\boldsymbol{\alpha}$ vector
\begin{equation}
    \begin{split}
        \boldsymbol{\alpha} &= ( 4.31338267, 4.3882788,  4.6578449, 4.67308374, 4.28873181, 4.26278301, \\
        &\mathrel{\phantom{=}} 4.73065817, 4.29330435, 4.44416548, 4.66625973, 4.26243104, 4.65881579, \\
        &\mathrel{\phantom{=}} 4.68086764, 4.44092086, 4.49639124, 4.55500032, 4.33389054, 4.38869161, \\
        &\mathrel{\phantom{=}} 4.57278526, 4.62717616, 4.62025928, 4.49780551, 4.46750298, 4.49561326, \\
        &\mathrel{\phantom{=}} 4.66902393, 4.60858869, 4.6027906, 4.40563641, 4.54198743, 4.49388045 ),
    \end{split}
    \label{eq:big-alpha-vector}
\end{equation}
with $\alpha_i\in(4.25,4.75)$.

\begin{adjustwidth}{-\extralength}{0cm}
\printendnotes[custom] 

\reftitle{References}

\PublishersNote{}
\end{adjustwidth}
\end{document}